\documentclass{LMCS}

\def\doi{8 (1:25) 2012}
\lmcsheading%
{\doi}
{1--31}
{}
{}
{Jan.~10, 2011}
{Mar.~16, 2012}
{}

\usepackage{hyperref, enumerate}
\usepackage{amsmath,amsfonts,amssymb}

\newcommand{\Z}{{\mathbb Z}}
\newcommand{\N}{{\mathbb N}}

\newcommand{\nL}{{\# L}}

\DeclareMathOperator{\comp}{-COMP}
\DeclareMathOperator{\ind}{-IND}

\newcommand{\LAP}{\textrm{LA{\sc p}}}
\newcommand{\LA}{\textrm{LA}}
\newcommand{\VpL}{V {\oplus} L}
\newcommand{\VnL}{V {\# L}}
\newcommand{\Pow}{{\operatorname{\it Pow}}}
\newcommand{\PowSeq}{{\operatorname{\it PowSeq}}}
\newcommand{\dm}{{\frac{\:.\:}{}}}
\newcommand{\row}{{\operatorname{r}}}
\newcommand{\ent}{{\operatorname{e}}}
\newcommand{\col}{{\operatorname{c}}}
\newcommand{\e}{{\operatorname{e}}}

\newcommand{\Parity}{{\operatorname{\it Parity}}}

\newcommand{\PAR}{{\operatorname{\it PAR}}}  
\newcommand{\p}{{\text{\sc p}}}
\newcommand{\strip}{{\operatorname{\it Strip}}}
\newcommand{\wrap}{{\operatorname{\it Wrap}}}
\newcommand{\isMatrixtwo}{{\operatorname{\it isMatrix}_2}}
\newcommand{\isMatrixz}{{\operatorname{\it isMatrix}_\Z}}

\newcommand\dotminus{\mathop{\mbox{$-^{\hspace{-.5em}\cdot}\,\,$}}}
\def\LtwoA{{\mathcal L}^2_A}
\def\calL{{\mathcal L}}
\def\calT{{\mathcal T}}
\def\ra{\rightarrow}
\def\lra{\leftrightarrow}
\def\comp{\mbox{-}\mathbf{COMP}}
\def\ind{\mbox{-}\mathbf{IND}}

\newcommand{\pairleft}{{\operatorname{\it left}}}
\newcommand{\pairright}{{\operatorname{\it right}}}

\newcommand{\parityL}{{\oplus L}}

\newcommand{\seq}{\operatorname{\it seq}}
\newcommand{\Pair}{\operatorname{\it Pair}}
\newcommand{\Row}{\operatorname{\it Row}}
\newcommand{\Rowtwo}{\operatorname{\it Row_2}}
\newcommand{\entry}{\operatorname{\it entry}}
\newcommand{\Powtwo}{\operatorname{\it Pow}_2}
\newcommand{\PowSeqtwo}{\operatorname{\it PowSeq}_2}
\newcommand{\PowSeqtwostar}{\operatorname{\it PowSeq}_2^\star}
\newcommand{\Prodtwo}{\operatorname{\it Prod}_2}
\newcommand{\IntDet}{\operatorname{\it IntDet}}

\newcommand{\numones}{\operatorname{\it numones}}
\newcommand{\bin}{\operatorname{\it bin}}

\newcommand{\intsize}{\operatorname{\it intsize}}
\newcommand{\PowZ}{{\operatorname{\it{ Pow}_{\Z}}}}

\newcommand{\PowSeqZ}{\operatorname{\it PowSeq}_\Z}

\newcommand{\ProdZ}{\operatorname{\it Prod}_\Z}
\newcommand{\Sum}{\operatorname{\it Sum}}

\begin{document}

\title{Formal Theories for Linear Algebra\rsuper*}

\author[S.~Cook]{Stephen Cook}
\address{University of Toronto
Department of Computer Science
Sandford Fleming Building
10 King's College Road
Toronto, Ontario M5S 3G4 
Canada}
\email{\{sacook,fontes\}@cs.toronto.edu}

\author[L.~Fontes]{Lila Fontes}
\address{\vskip-6 pt}

\titlecomment{{\lsuper*}A preliminary version of this work appeared as
  \cite{CookFo10}.}


\begin{abstract}
We introduce two-sorted theories in the style of Cook and Nguyen
for the complexity classes $\parityL$ and $DET$, whose complete problems
include determinants over $\Z_2$ and $\Z$, respectively.
We then describe interpretations of Soltys' linear algebra theory $\LAP$
over arbitrary integral domains, into each of our new theories.
The result shows equivalences of standard theorems of linear algebra
over $\Z_2$ and $\Z$ can be proved in the corresponding theory, but
leaves open the interesting question of whether the theorems
themselves can be proved.
\end{abstract}

\maketitle


\section{Introduction} \label{s:intro}

This paper is a contribution to bounded reverse mathematics
\cite{Nguyen08,CookNg10}, that part of proof complexity concerned with
determining the computational complexity of concepts needed to
prove theorems of interest in computer science.  We are specifically
interested in theorems of linear algebra over finite fields and
the integers.  The relevant complexity classes for each case
have been well-studied in the computational complexity
literature.  The classes are $\parityL$ and $DET$, associated with linear
algebra over $\Z_2$ and $\Z$, respectively.  We introduce formal
theories $V\parityL$ and $V\#L$ for $\parityL$ and $DET$, each
intended to capture reasoning in the corresponding class.
Each theory allows induction over any relation in the
associated complexity class, and the functions
definable in each theory are exactly the functions in the class. 
In particular determinants and coefficients of the
characteristic polynomial of a matrix can be defined.

To study the question of which results from linear
algebra can be proved in the theories we take advantage of
Soltys's theory $\LAP$ \cite{Soltys01,SoltysCo04} for formalizing linear
algebra over an arbitrary field or integral domain.
We present two interpretations of $\LAP$:
one into $V\parityL$ and one into $V\#L$.
Both interpretations translate theorems of $\LAP$ to theorems in
the corresponding theory, but the meaning of the theorems differs
in the two translations since the ring elements range over
$\Z_2$ in one and over $\Z$ in the other.
From these interpretations and results in \cite{Soltys01,SoltysCo04}
we show that the theories prove some interesting properties of
determinants, but leave open the question of whether the proofs
of some basic theorems such as the Caley-Hamilton Theorem can be
formalized in the theories.  We also leave open the 
question of whether the theories prove simple matrix identities
studied in \cite{Soltys01,SoltysCo04}, such as $AB = I \ra BA = I$.
An affirmative answer would shed light on interesting questions
in propositional proof complexity concerning the lengths of
proofs required in various proof systems to prove tautology
families corresponding to the identities.

\subsection{The complexity classes}\label{s:Cclasses}

Complete problems for the classes $\parityL$ and $DET$ include standard
computational problems of linear algebra over their respective rings
$\Z_2$ and $\Z$, such as computing
determinants, matrix powers, and coefficients of the
characteristic polynomial of a matrix.  (Recently
\cite{BravermanKuRo09} proved that for each $k\ge 1$, computing the
permanent mod $2^k$ of an integer matrix is in $\parityL$, and hence
complete.)  The classes satisfy the inclusions
\begin{equation} \label{e:PL}
   \begin{array}{ll}
    & AC^0 \subset L \subseteq \parityL \subseteq DET \subseteq NC^2
\subseteq P    \\  
    & AC^0 \subset L \subseteq NL \subseteq DET \subseteq NC^2
    \end{array}
\end{equation}
(ignoring the distinction between function and language
classes) where $L$ and $NL$ are the problems accepted in deterministic
and nondeterministic log space, respectively.  It is not known whether
$\parityL$ and $NL$ are comparable.  (In fact no one has been able
to disprove the unlikely possibility that all of the above classes
except $AC^0$ coincide.)

The simplest way of defining the classes $\parityL$ and $DET$ is using
uniform $AC^0$ reductions:  Let $AC^0(f)$ be the set of functions
computable by a uniform family of polynomial size constant depth
circuits with oracle access to $f$.  Then 
$\parityL = AC^0(det_2)$ and $DET = AC^0(det)$, where $det_2$ and 
$det$ are the determinant functions for matrices over $\Z_2$ and $\Z$ 
respectively.  For the case of $\Z$, integer entries of a matrix
are presented in binary.

The usual definitions of these classes involve counting the number of
accepting computations of nondeterministic log space Turing machines.
Thus $\#L$ is the class of functions $f$ such that for some
nondeterministic log space Turing machine $M$, $f(x)$ is the number of
accepting computations of $M$ on input $x$.  Then the sets in
$\parityL$ are those of the form $\{x \mid f(x) \bmod 2 = 1\}$ for
some $f$ in $\#L$.  It turns out that $AC^0(det) = AC^0(\#L)$, and
$AC^0(det_2) = AC^0(\parityL)=\parityL$
\cite{AllenderOg96,BuntrockDaHeMe92}.

$DET$ can also be characterized as the $\#L$ hierarchy $\#LH$.
This is defined as follows:
$\#LH_1 = \#L$, and for $i\ge 1$, $\#LH_{i+1} = \#L^{\#LH_i}$.
(The exponent $\#LH_i$ indicates that a function from this class is
allowed to be an oracle for the log space machine whose accepting
computations are being counted).   Then \cite{AllenderOg96} shows
\begin{equation}\label{e:DET=LH}
DET = \#LH = \bigcup_i \#LH_i
\end{equation}

We should clarify that our definition of $DET = AC^0(det)$ here
differs from that given
in \cite{Cook85}, where $DET$ is defined to be $NC^1(det)$, the
closure of $\{det\}$ under the more general $NC^1$ reductions.
Allender proved (see the Appendix to \cite{Allender04}) that if
$AC^0(det) = NC^1(det)$ then the $\#L$ hierarchy collapses to some
finite level $\#LH_i$, something that is not known to be true. 
However the present first author wrote \cite{Cook85} before uniform
$AC^0$ reductions
had been studied, and now believes that $AC^0$
reductions are the natural ones to use in studying small complexity
classes.   Evidence for this is that (\ref{e:DET=LH})
holds when $DET = AC^0(det)$ as we now define $DET$, and
does not hold under the old definition (assuming the $\#L$ hierarchy
is strict).

The above inclusions (\ref{e:PL})
compare the class of functions $DET$ with classes
of relations.  Here and elsewhere we sometimes do not explicitly
distinguish between a function class $FC$ and the corresponding
relation class $C$, using the following standard correspondence (which
applies to all classes we consider): A relation is in $C$ iff its
characteristic function is in $FC$, and a function is in $FC$ iff it
is polynomially bounded and its bit graph is in $C$.  (The
bit graph of a function $F:\{0,1\}^\ast \ra \{0,1\}^\ast$ is the
relation $B_F(i,X)$ which holds iff the $i$th bit of $F(X)$ is 1.)

\subsection{The theories \texorpdfstring{$\VpL$}{VparityL} and \texorpdfstring{$\VnL$}{VnumberL}}

To construct formal theories for the classes $\parityL$ and $DET$
we follow the framework laid out in Chapter 9 of the monograph
\cite{CookNg10} of Cook and Nguyen for defining theories for certain
complexity classes between $AC^0$ and $P$.
All of these theories share a common two-sorted (number and string)
vocabulary $\LtwoA$ (see Equation \ref{e:LtwoA}).
The intention is that the number sort ranges over $\N$ and the string 
sort ranges over bit strings (more precisely, finite subsets of $\N$).
The strings are intended to be inputs to the machine or circuit
defining a member of the complexity class, and the numbers are
used to index bits in the strings.
Each theory $VC$ for a class $C$ extends the
finitely-axiomatized base
theory $V^0$ for $AC^0$ by addition of a single axiom stating the
existence of a solution to a complete problem for $C$.  General
techniques are presented for defining a universally-axiomatized
conservative extension $\overline{VC}$ of $VC$ which has function
symbols and defining axioms for each function in $FC$, and
$\overline{VC}$ admits induction on open formulas in this enriched
vocabulary.  It follows from the Herbrand Theorem that the
provably-total functions in $\overline{VC}$ (and hence in $VC$) 
are precisely the functions in $FC$.

Chapter 9 (with earlier chapters) of \cite{CookNg10} explicitly
defines theories for the following classes:
\begin{equation}\label{e:Tclasses}
  AC^0 \subset AC^0(2) \subset TC^0\subseteq NC^1 
  \subseteq L \subseteq NL \subseteq NC \subseteq P
\end{equation}
These classes are defined briefly as follows.
A problem in $AC^0$ is solved by a uniform family of polynomial size
constant depth Boolean circuits with unbounded fanin AND and OR gates.
$AC^0(2)$ properly extends $AC^0$ by also
allowing unbounded fanin parity gates
(determining whether the inputs have an odd number of 1's) in its circuits.
$TC^0$ allows majority gates rather than parity gates in its circuits
(and has binary integer multiplication
as a complete problem).  $NC^1$ circuits restrict all
Boolean gates to fanin two,
but the circuits are allowed to have logarithmic depth. 
Problems in $L$ and $NL$ are solved respectively by deterministic and
nondeterministic log space Turing machines.
$NC$ is defined like $NC^1$,
but the circuits can have polylogarithmic depth (and polynomial size).
Problems in $P$ are solved by polynomial time Turing machines.

Our new theories $\VpL$ and $\VnL$ for $\parityL$ and $DET$
extend the base theory $V^0$ for $AC^0$ by adding axioms
stating the existence of powers $A^k$ of matrices $A$ over
$\Z_2$ and $\Z$, respectively.  Here $k$ is presented in unary,
but for the case of $\VnL$ integer entries for $A$ are presented
in binary.
(Matrix powering is a complete problem for these classes).
Here there is a technical difficulty of
how to nicely state these axioms, since neither the parity function
(needed to define matrix multiplication over $\Z_2$) nor integer
product and multiple summation (needed to define matrix multiplication
over $\Z$) are $AC^0$ functions, and hence neither
is definable in the base theory $V^0$.  
We solve this by basing an initial version of $\VpL$ on the theory
$V^0(2)$ for $AC^0(2)$ (which contains the parity function) and basing
an initial version of $\VnL$ on the
theory $VTC^0$ for $TC^0$ (which contains integer product). 
We then use results
from \cite{CookNg10} to translate the axioms for the initial versions
to the language of the base theory
$V^0$, to obtain the actual theories $\VpL$ and $\VnL$.  We show that
the resulting theories satisfy the requirements
of Chapter 9 (existence of ``aggregate functions'')
that allow the existence of the nice universal conservative extensions
$\overline{\VpL}$ and $\overline{\VnL}$ of $\VpL$ and $\VnL$.
Using general results from Chapter 9 of \cite{CookNg10} we show
the following (see Theorems \ref{p:bartwo} and \ref{p:barZ} for more
formal statements). 

\begin{thm}\label{t:fundThm}
The provably total functions of $\VpL$ and $\overline{\VpL}$
(resp. $\VnL$ and $\overline{\VnL}$) are exactly the functions
of the class $\parityL$ (resp. $DET$).  Further
$\overline{\VpL}$ (resp. $\overline{\VnL}$) proves
the induction scheme for $\Sigma_0^B(\mathcal{L}_{F\parityL})$ formulas
(resp. $\Sigma_0^B(\mathcal{L}_{F\nL})$ formulas.
\end{thm}

The last sentence means in effect that $\VpL$ and $\VnL$ prove the
induction schemes for formulas expressing concepts in their corresponding
complexity classes.

The new theories mesh nicely with the theories for the complexity
classes in (\ref{e:Tclasses}).  In particular, we have
\begin{equation}\label{e:mesh}
  V^0 \subset V^0(2) \subset VTC^0\subseteq VNC^1 \subseteq VL \subseteq   
 V\parityL \subseteq V\#L \subseteq VNC \subseteq VP
\end{equation}

We also have $VL \subseteq VNL \subseteq V\#L$.  We do not know whether
$VNL$ and $V\parityL$ are comparable, because we do not know whether
$NL$ and $\parityL$ are comparable.

\subsection{The interpretations}
Next we study the question of which results from linear
algebra can be proved in the theories.  As mentioned above,
we take advantage of
Soltys's theory $\LAP$ \cite{Soltys01,SoltysCo04} for
formalizing results from linear
algebra over an arbitrary field or integral domain.
We present two interpretations of $\LAP$:
one into $V\parityL$ and one into $V\#L$.  
Both interpretations translate theorems of $\LAP$ to theorems in
the corresponding theory, but the meaning of the theorems differs
in the two translations since the ring elements range over
$\Z_2$ in one and over $\Z$ in the other.



$\LAP$ defines matrix powering, and uses this definition and Berkowitz's 
algorithm
\cite{Berkowitz84} to define several functions of matrices, including determinant,
adjoint, and characteristic polynomial.   The following standard
principles of linear algebra are discussed:
\begin{enumerate}[(i)]
\item
The Cayley-Hamilton Theorem (a matrix satisfies its characteristic
polynomial).
\item
The axiomatic definition of the determinant (the function $det(A)$ is
characterized by the properties that it is
multilinear and alternating in the rows and columns of $A$,
and $det(I)=1$).
\item
The co-factor expansion of the determinant.
\end{enumerate}

Although it remains open whether $\LAP$ can prove any of these, a major
result from \cite{Soltys01,SoltysCo04} is that $\LAP$ proves their pairwise equivalence.
As a result of this and our interpretations (Theorems \ref{t:provability}
and \ref{t:NumProvability}) we have the following.

\begin{thm}\label{t:equiv}
$V\parityL$ proves the equivalence of (i), (ii), and (iii) over
the ring $\Z_2$, and $V\#L$ proves their equivalence over $\Z$.
\end{thm}

An intriguing possibility is that either
$V\parityL$ or $V\#L$ could use special properties of $\Z_2$ or
$\Z$ to prove its version of the principles, but
$\LAP$ cannot prove them (for all integral domains or fields).
For example there is a dynamic programming algorithm involving
combinatorial graph properties
(see the concluding Section \ref{s:conclusion})
whose correctness for $\Z$ might be provable in $V\#L$ using
combinatorial reasoning with concepts from $\#L$ which
are not available in $\LAP$.

\cite{Soltys01,SoltysCo04} also present the so-called
{\em hard matrix identities}:
\begin{defi}\label{d:hardM}
The {\em hard matrix identities} are
\begin{equation}\label{e:hard}
\begin{split}
&  AB  =  I, AC = I \ra B=C\\
&  AB = I, AC = 0  \ra C = 0 \\
&  AB = I \ra BA = I \\
& AB = I  \ra A^tB^t = I
\end{split}
\end{equation}
where $A,B,C$ are square matrices of the same dimensions,
and $A^t$ is the transpose of $A$.
\end{defi}
Again it is open whether $\LAP$ proves these identities, but
$\LAP$ does prove that they follow from
any of the principles mentioned in Theorem \ref{t:equiv} above.
The next result follows from this and our interpretations
(Theorems \ref{t:provability} and \ref{t:NumProvability}).
\begin{thm}\label{t:hardM}
$V\parityL$ proves that (\ref{e:hard})
over the ring $\Z_2$ follows from any of the
three principles mentioned in Theorem \ref{t:equiv}.   The same is
true for $V\#L$ over the ring $\Z$.   
\end{thm}

\cite{Soltys01,SoltysCo04} introduce an extension $\forall\LAP$ of
$\LAP$, which includes an induction rule that applies to
formulas with bounded universally
quantified matrix variables, and show that the three principles
mentioned in Theorem \ref{t:equiv} and the four matrix identities
are all provable in $\forall\LAP$.   These papers claim
that these proofs in $\forall\LAP$ translate into proofs in the
theory $V^1$ for polynomial time ($V^1$ extends $VP$ in equation
(\ref{e:mesh})) when the underlying ring is finite or $\mathbb{Q}$.
However Je{\v r}{\'a}bek \cite{Jer05} (page 44) points out that 
for infinite rings this is not true, because the definition given
of bounded universal matrix quantifiers only bounds the number of
rows and columns, and not the size of the entries.  To fix this,
\cite{Jer05} defines a subsystem $\forall\LAP^-$ of $\forall\LAP$, with
properly defined bounded universal matrix quantifiers, which still
proves the three principles mentioned in Theorem \ref{t:equiv} and the
four matrix identities, and shows that these proofs translate
into proofs in $V^1$ when the underlying ring is finite or $\mathbb{Q}$
(and hence also $\Z$).  Since $V^1$ is conservative over $VP$
for universal theorems involving polynomial time functions we have
the following result.
\begin{prop}\label{t:polytime}
\cite{Soltys01,SoltysCo04,Jer05}
The theory $VP$ proves the three principles (i), (ii), (iii)
and the matrix identity (\ref{e:hard})
for both the rings $\Z_2$ and $\Z$.
\end{prop}


\section{Two-Sorted Theories}\label{s:theories}

We start by reviewing the two-sorted logic used here and in
\cite{CookNg10}.
We have {\em number} variables $x,y,z,\ldots $
whose intended values are numbers (in $\N$), and {\em string}
variables $X,Y,Z,\ldots $ whose intended values are finite sets
of numbers.  We think of the finite sets as binary strings giving
the characteristic vectors of the sets.  For example the string
corresponding to the set $\{0,3,4\}$ is $10011$.

All our two-sorted theories include the basic vocabulary $\LtwoA$,
which extends the first-order vocabulary of Peano Arithmetic as
follows:
\begin{equation}\label{e:LtwoA}
\LtwoA = [0,1,+,\cdot, | \ |, \in, \le,=_1,=_2]
\end{equation}
The symbols $0,1,+,\cdot$ are intended to take their usual
meanings on $\N$.  Here $| \ |$ is a function from strings to
numbers, and the intended meaning of $|X|$ is 1 plus the largest
element of $X$, or 0 if $X$ is empty.  (If $X = \{0,3,4\}$ then
$|X| = 5$.)  The binary predicate $\in$ is intended to denote
set membership.  We often write $X(t)$ for $t\in X$ (think that bit
number $t$ of the string $X$ is 1).  The equality predicates
$=_1$ and $=_2$ are for numbers and strings, respectively.
We will write = for both, since the missing subscript will be clear
from the context.

Number terms (such as $x + ((|X|+1)\cdot |Y|)$)
are built from variables and function symbols as usual.
The only string terms based on $\LtwoA$ are string variables
$X,Y,Z,\ldots$, but
when we extend $\LtwoA$ by adding string-valued functions, other
string terms will be built as usual.  Formulas are built from atomic
formulas (e.g. $t=u, t \le u, X(t), X=Y$) using $\wedge,\vee,\neg$
and $\exists x,\forall x,\exists X, \forall X$.

Bounded quantifiers are defined as usual, except bounds on string
quantifiers refer to the length of the string.
For example $\exists X{\le}t \; \varphi$ stands for
$\exists X (|X|{\le}t \, \wedge \, \varphi)$.

We define two important syntactic classes of formulas.

\begin{defi}\label{d:Sigforms}
$\Sigma^B_0$ is the class of $\LtwoA$ formulas with no
string quantifiers, and only bounded number quantifiers.
$\Sigma^B_1$ formulas are those of the form
$\exists \vec{X}{\le} \vec{t} \; \varphi$, where $\varphi$ is
in $\Sigma^B_0$ and the prefix of bounded quantifiers may be empty.
\end{defi}

Notice our nonstandard requirement that the string quantifiers
in $\Sigma^B_1$ formulas must be in front.

We also consider two-sorted vocabularies $\calL \supseteq \LtwoA$
which extend $\LtwoA$ by possibly adding predicate symbols $P,Q,R,\ldots$
and function symbols $f,g,h,\ldots$ and $F,G,H,\ldots$.
Here $f,g,h,\ldots$
are {\em number functions} and are intended to take values in $\N$,
and $F,G,H,\ldots$ are {\em string functions} and are intended to
take string values.  Each predicate or function symbol has a
specified arity $(n,m)$ indicating that it takes $n$ number arguments
and $m$ string arguments.  Number arguments are written before
string arguments, as in
\begin{equation}\label{e:functions}
 f(x_1,\ldots,x_n,X_1,\ldots,X_m)  \qquad
 F(x_1,\ldots,x_n,X_1,\ldots,X_m)
\end{equation}
The formula classes $\Sigma^B_0(\calL)$ and $\Sigma^B_1(\calL)$ are
defined in the same way as $\Sigma^B_0$ and $\Sigma^B_1$, but allow
function and relation symbols from $\calL$ in addition to
$\LtwoA$.

\subsection{Two-sorted complexity classes} \label{s:two-classes}

In standard complexity theory an element of a  complexity class is
either a set of binary strings or a function
$f:\{0,1\}^* \ra \{0,1\}^*$.  In our two-sorted point of view (Chapter 4
of \cite{CookNg10}) it is convenient to replace a set of strings by a relation
$R(\vec{x},\vec{X})$ of any arity $(n,m)$, and functions
are generalized to allow both number functions and string functions
as in (\ref{e:functions}).  Each standard complexity class, including
those in (\ref{e:Tclasses}) and $\parityL$ and $\#L$, is defined
either in terms of Turing machines or circuit families.  These
definitions naturally extend to two-sorted versions by representing
strings (as inputs to machines or circuits) in a straightforward way
as binary strings, but by representing
numbers using unary notation.  This interpretation of numbers is a
convenience, and is justified by our intention that numbers are
`small' and are used to index strings and measure their length.

For example, the (two-sorted) complexity class P (resp. NP) is the set of
relations $R(\vec{x},\vec{X})$ recognized by polynomial time
(resp. nondeterministic polynomial time)
Turing machines, with the above input conventions.  
Thus the relation $Prime_1(x)$ ($x$ is a prime number) is trivially
in P since there are at most $x$ possible divisors to test, and the
testing can obviously be done in time polynomial in $x$.
The relation $Prime_2(X)$ (the number whose binary notation is $X$
is prime) is also in P, but this is a major result \cite{AKS04},
since the testing must be done in time polynomial in the length
$|X|$ of $X$.

The class (uniform) $AC^0$ can be defined in terms of uniform
polynomial size constant depth circuit families, but
it has a nice characterization as those sets recognized by an
alternating Turing machine (ATM) in log time with a constant number
of alternations.   More useful for us, \cite{Immerman99} showed that
an element of $AC^0$ can be described as an element of $FO$,
namely the set of finite models of some first-order formula with a
certain vocabulary.  From this and the ATM definition
of two-sorted $AC^0$, we have the following
important results relating syntax and semantics.

\begin{prop}\label{p:repThm}
[Representation Theorems]
(IV.3.6 and IV.3.7 of \cite{CookNg10})
A relation $P(\vec{x},\vec{X})$ is in $AC^0$ (respectively $NP$)
iff it is represented by
some $\Sigma^B_0$-formula (respectively $\Sigma^B_1$-formula)
$\varphi(\vec{x},\vec{X})$.
\end{prop}

For example the relation $\operatorname{\it PAL}(X)$ ($X$ is a palindrome) is an
$AC^0$ relation because the $\Sigma^B_0$-formula
$ \forall x,y{ <} |X| \; (x+y+1 = |X| \supset (X(x)\leftrightarrow X(y)))$
represents it.

\subsection{Special functions} \label{s:specialfns}

There are several conventional number and string functions
in $FAC^0$ used to encode and retrieve information from bit-strings.
Their usage is crucial to several technical proofs below, so we here
provide definitions of common functions from \cite{CookNg10} as well
as extensions useful for our own purposes.

The pairing function $ \langle x , y \rangle  = (x+y)(x+y+1) + 2y $
allows easy $2$-dimensional indexing in strings.  It can be extended
to $k$-arity.
\[
 \langle x_1, x_2, \ldots, x_k \rangle =
\langle  x_1, \langle x_2, \ldots, x_k\rangle\rangle
\label{eq:pairing}
\]
We construct the unary relation $\Pair(z)$ to be true only of
numbers in the range of the pairing function.
The functions $\pairleft(\langle x,y\rangle)=x$ and
$\pairright(\langle x,y \rangle)=y$ reverse the pairing function, and
are defined to be $0$ on numbers $z$ where $\Pair(z)$ is false.

For example a $k$-dimensional bit array can be encoded by a string $X$,
with
bits recovered using the pairing function.  By convention, we write:
\[ X(x_1,\ldots,x_k) = X(\langle x_1,\ldots, x_k\rangle) \]

For $2$-dimensional arrays, the $x^\textrm{th}$ row can be recovered
using the $\Row$ function, which is bit-defined:
\[
\Row(x,Z)(i) \leftrightarrow i < |Z| \wedge Z(x,i)
\]
For notational convenience, we write $\Row(x,Z) = Z^{[x]}$.
Thus a $1$-dimensional array of $j$ strings $X_1, \ldots, X_j$
can be encoded in a single bit-string $Z$, where $X_i = Z^{[i]}$.

It will be useful when dealing with integers (section \ref{s:VnL}) to
be able to encode $2$-dimensional arrays of strings $X_{i,j}$ as a
single string $Z$.  To do so, we first encode each row of the
$2$-dimensional matrix of strings as a $1$-dimensional array of
strings $Y_i$ so that $Y_i^{[j]} = X_{i,j}$.  Then we encode a
$1$-dimensional array of \emph{those} arrays: $Z^{[i]}=Y_i$.  The
$x,y^\textrm{th}$ strings in this $2$-dimensional array $Z$ is
recovered with the $\Rowtwo$ function: $\Rowtwo(x,y,Z) =
Z^{[x][y]}$.
\begin{equation} \label{eq:rowtwo}
\Rowtwo(x,y,Z)(i) \leftrightarrow i<|Z| \wedge \Row(x,Z)(y,i)
\end{equation}

The $\Row$ and $\Rowtwo$ functions allow a bit-string $Z$ to encode a
list or matrix of other bit-strings.  We have analogous functions
$\seq$ and $\entry$ that enable string $Z$ to encode a list or matrix
of numbers (respectively).  Thus the numbers $y_0$, $y_1$, $y_2$,
\ldots encoded in $Z$ can be recovered $y_i = \seq(i,Z) = (Z)^i$ by
convention.
\begin{align} 
& y = \seq(x,Z) \leftrightarrow \nonumber \\
& (y< |Z| \wedge Z(x,y) \wedge \forall z<y, \neg Z(x,z)) \vee
(\forall z<|Z|, \neg Z(x,z) \wedge y=|Z|) 
\end{align}
Numbers $y_{i,j}$ are encoded in $Z$ by first arranging them into rows
$Y_i$ where $(Y_i)^j = \seq(j,Y_i)$. Then $Y_i = Z^{[i]}$.
The numbers are retrieved using the $\entry$ function: $y_{i,j} =
\entry(i,j,Z)$.
\[
\entry(i,j,Z)=y \leftrightarrow (Z^{[i]})^j = y
\]

\subsection{The classes \texorpdfstring{$\parityL$}{parityL} and \texorpdfstring{$DET$}{DET}}\label{s:newClasses}

\begin{defi}\label{d:parityL}
$\parityL$ is the set of relations
$R(\vec{x},\vec{X})$ such that
there is a nondeterministic log space Turing machine $M$ such that
$M$ with input $\vec{x},\vec{X}$ (represented as described in
Section \ref{s:two-classes}) has
an odd number of accepting computations iff $R(\vec{x},\vec{X})$ holds.

$\#L$ is the set of string functions $F(\vec{x},\vec{X})$ such that
there exists nondeterministic log space Turing machine $M$,
and $F(\vec{x},\vec{X})$ is the number (in
binary) of accepting computations of $M$ with input $\vec{x},\vec{X}$.
\end{defi}

In the above definition we restrict attention to log space Turing
machines that halt for all computations on all inputs.
We think of the binary string $F(\vec x, \vec X)$ as a number, as
given by the following definition.
\begin{defi} \label{def:bin-fn}
A string $X$ represents the number $\bin(X)$ if
\[
    \bin(X)  = \sum_i 2^i X(i) 
\]
where we treat the predicate $X(i)$ as a 0-1 valued function.
\end{defi}

We now formalize the correspondence (mentioned in Section \ref{s:intro})
between a complexity class $C$ of relations and a complexity class
$FC$ of functions.
A number function $f(\vec{x},\vec{X})$ (respectively string function
$F(\vec{x},\vec{X})$) is $p$-bounded if there is a polynomial
$g(\vec{x},\vec{y})$ such that
$f(\vec{x},\vec{X}) \le g(\vec{x}, |\vec{X}|)$ (respectively
$|F(\vec{x},\vec{X})| \le g(\vec{x}, |\vec{X}|)$).  The
{\em bit graph} of a string function $F$ is the relation $B_F$
defined by $B_F(i,\vec{x},\vec{X}) \lra F(\vec{x},\vec{X})(i)$.
%
\begin{defi}\label{d:FC}
If $C$ is a class of (two-sorted) relations then $FC$ denotes
the corresponding class of functions, where $f$ (respectively $F$)
is in $FC$ iff it is $p$-bounded and its graph (respectively bit graph)
is in $C$.  If $C$ is a two-sorted complexity class of functions,
then $RC$ consists of all relations whose characteristic functions
are in $C$.
\end{defi}

In general when we refer to a complexity class such as $AC^0$
or $P$ we refer to the relations in the class, and sometimes
also to the functions in the corresponding function classes
$FAC^0$ and $FP$.

We will consider two-sorted vocabularies $\calL$ which extend
$\LtwoA$, and in all cases each function and relation symbol
in $\calL$ has a specific intended interpretation in our standard
two-sorted model (the two universes being $\N$ and
the set of finite subsets of $\N$).  Thus we can make sense of
both syntactic and semantic statements about $\calL$.

If $\calL$ is a two-sorted vocabulary, then $f$ (respectively $F$)
is $\Sigma^B_0$-definable from $\calL$ if it is $p$-bounded
and its graph (respectively bit graph) is represented by a
formula in $\Sigma^B_0(\calL)$.  In this case $f$ (or $F$) can
be computed by a uniform polynomial size family of bounded-depth
circuits with oracle access to the functions and predicates in $\calL$.

The following definition is from page 269 of \cite{CookNg10}.

\begin{defi}[$AC^0$-reducibility] \label{d:AC0reduc}
(IX.1.1 in \cite{CookNg10})
A string function $F$ (respectively, a number function $f$) is
$AC^0$-reducible to $\mathcal{L}$ if there is a sequence of string
functions $F_1$, \ldots, $F_n$, ($n\geq 0$) such that
\[ F_i \text{ is } \Sigma_0^B\text{-definable from }
  \mathcal{L} \cup \{ F_1, \ldots, F_{i-1} \}
  \text{ for }
  i = 1, \ldots, n;
\]
and $F$ (resp. $f$) is $\Sigma_0^B$-definable from
$\mathcal{L} \cup \{ F_1, \ldots, F_{n}\}$. 
A relation $R$ is $AC^0$-reducible to
$\mathcal{L}$ if there is a sequence of string functions
$F_1$, \ldots, $F_n$ as above, and $R$ is represented by a
$\Sigma_0^B(   \mathcal{L} \cup \{ F_1, \ldots, F_{n}\})$-formula.

$AC^0(\calL)$ (the $AC^0$ closure of $\calL$) denotes the closure
of $\calL$ under $AC^0$-reducibility. 
\end{defi}

 It is easy to see
from the above definitions and Proposition \ref{p:repThm} that
$AC^0(\LtwoA)$ is the set of all $AC^0$ functions and relations.

From Lemma 6 in \cite{BuntrockDaHeMe92}  we know that $\parityL$ is
closed under $NC^1$ reductions, and hence also under the weaker
$AC^0$ reductions, so $AC^0(\parityL) = \parityL$.
However we do not know whether $\#L$ is closed under $AC^0$ reductions
\cite{AllenderOg96}.   Since our theories can only characterize
classes closed under $AC^0$ reductions, we use the class $DET$
instead of $\#L$.

\begin{defi}\label{d:DET}
$DET = AC^0(\#L)$
\end{defi}

Matrix powering is central to our theories for $\parityL$ and $DET$.
So we need to define this for each of the two classes as a two-sorted
function.  For $\parityL$ we use the truth values $\{\operatorname{\it
  false}, \operatorname{\it true}\}$
to represent the elements
$\{0,1\}$ of $\Z_2$, and we represent a matrix over $\Z_2$ with a
string $X$.  For $DET$ we represent integers with binary notation
by bit strings (using Definition \ref{def:bin-fn}), and
we represent a matrix over $\Z$  by an array of strings
(see Section \ref{s:specialfns}).
We number rows and columns starting with 0, so if $X$ is
an $n\times n$ matrix then $0\le i,j < n$ for all entries $X(i,j)$.

\begin{defi}[Matrix Powering] \label{d:Pow}
Let $X$ be a string representing an $n\times n$ matrix over $\Z_2$
(resp. $\Z$).
Then the string function $\Powtwo(n,k,X)$ (resp. $\PowZ(n,k,X)$)
has output $X^k$, the string
representing the $k^\textrm{th}$ power of the same
matrix.
\end{defi}

\begin{prop}\label{p:completeness}
$\parityL = AC^0(\Powtwo)$ and $DET = AC^0(\PowZ)$.
\end{prop}

\proof
For $DET$, by Definition \ref{d:DET} it suffices to show 
$AC^0(\PowZ) = AC^0(\#L)$, and this is proved in \cite{Fontes09}.
A proof can also be extracted from the earlier literature as follows.
Let $\IntDet(n,X)$ be the function that
returns the determinant of the $n\times n$ integer matrix $X$.
By Berkowitz's algorithm \cite{Berkowitz84,SoltysCo04},
$\IntDet$ is $AC^0$ reducible to $\PowZ$, and by a standard reduction
(see for example the proof of Proposition 5.2 in \cite{Cook85})
$\PowZ$ is $AC^0$ reducible to $\IntDet$.   Hence it remains to
show that $AC^0(\IntDet) = AC^0(\#L)$, and this is stated as Corollary 21
in \cite{AllenderOg96}.  A key idea is to show that the problem of
counting the number of paths of length at most a parameter $p$
from nodes $s$ to $t$ in a directed graph is complete for $\#L$.

The fact that $\parityL = AC^0(\Powtwo)$ follows from Theorem 10
in \cite{BuntrockDaHeMe92}.  (That theorem is stated in terms of
$NC^1$ reductions, but the needed reductions are easily shown to
be $AC^0$.)
\qed

Although $\PowZ(n,k,X)$ allows the entries of the matrix $X$ to
be arbitrary binary integers, it turns out that the function is
still complete for $DET$ when the argument $X$ is restricted to
matrices with entries
in ${0,1}$ (see for example the proof of Lemma 16 in \cite{Fontes09}).

\subsection{The Theories \texorpdfstring{$V^0$}{V0}, \texorpdfstring{$V^0(2)$}{V02}, and \texorpdfstring{$VTC^0$}{VTC0}}\label{s:threeV}

The theory $V^0$ for $AC^0$ is the basis for every two-sorted theory
considered here and in \cite{CookNg10}.  It has the two-sorted vocabulary
$\LtwoA$, and is
axiomatized by the set 2-BASIC (Figure \ref{f:basic})
of axioms consisting of 15 $\Sigma^B_0$
formulas expressing basic properties of the symbols of $\LtwoA$,
together with the following $\Sigma^B_0$ comprehension scheme.
\[
\Sigma^B_0\comp: \ \   \exists X {\le} y \; \forall z{<}y \; (X(z)\lra \varphi(z))
\]
Here $\varphi(z)$ is any $\Sigma^B_0$ formula with no free occurrence
of $X$.

\begin{figure}[h]
\begin{tabular}{ll}
{\bf B}{1.} $x+1\neq 0$                         & {\bf B}{2.} $x+1 = y+1\supset x=y$ \\
{\bf B}{3.} $x+0 = x$                           & {\bf B}{4.} $x+(y+1) = (x+y) + 1$ \\
{\bf B}{5.} $x\cdot 0 = 0$                      & {\bf B}{6.} $x\cdot (y+1) = (x\cdot y) + x$ \\
{\bf B}{7.} $(x\le y \wedge y\le x)\supset x = y$                & {\bf B}{8.} $x \le x + y$\\
{\bf B}{9.} $0 \le x$                           & {\bf B}{10.} $x\le y \vee y\le x$\\
{\bf B}{11.} $x\le y \lra x< y+1$               & {\bf B}{12.} $x\neq 0 \supset \exists y\le x(y+1 = x)$\\

{\bf L}{1.} $X(y) \supset y < |X|$                & {\bf L}{2.}  $y+1 = |X| \supset X(y)$ \\
\multicolumn{2}{l}
{{\bf SE}{.} $\big(|X| = |Y|\wedge\forall i<|X|(X(i)\lra Y(i))\big)\ \supset\ X = Y$}
\end{tabular}
\caption{2-BASIC axioms}\label{f:basic}
\end{figure}

$V^0$ has no explicit induction axiom, but nevertheless the induction
scheme
\[
   \Sigma^B_0 \ind:  \ \big(\varphi(0) \wedge \forall x(\varphi(x) \supset
\varphi(x+1))\big)  \supset \forall z\varphi(z)
\]
for $\Sigma^B_0$ formulas $\varphi(x)$ is provable in $V^0$, using
$\Sigma^B_0\comp$ and the fact that $|X|$ produces the maximum
element of the set $X$.

\begin{defi}\label{d:SigDefine}
A string function $F(\vec{x},\vec{X})$ is $\Sigma^B_1$-{\em definable}
(or {\em provably total})
in a two-sorted theory $\calT$ if there is a $\Sigma^B_1$ formula
$\varphi(\vec{x},\vec{X},Y)$ representing the graph
$Y=F(\vec{x},\vec{X})$ of $F$ such that
$\calT \vdash \forall \vec{x} \; \forall \vec{X} \; \exists! Y \;
\varphi(\vec{x},\vec{X},Y)$.
Similarly for a number function $f(\vec{x},\vec{X})$.
\end{defi}

It is shown in Chapter 5 of \cite{CookNg10} that $V^0$ is finitely
axiomatizable, and the $\Sigma^B_1$-definable functions in $V^0$
comprise the class $FAC^0$ (see Definition \ref{d:FC}).

The definition in \cite{CookNg10} of the theory $V^0(2)$ for the class
$AC^0(2)$ is based on $V^0$ and an axiom showing the definability of
the function $\Parity(x,Y)$.  If $Z = \Parity(x,Y)$ then $Z(z)$ holds
iff $1\le z \le x$ and there is an odd number of ones in $Y(0) Y(1)
\ldots Y(z\dotminus 1)$.  The graph of $\Parity$ is defined by the
following $\Sigma^B_0$ formula:
\[
 \delta_{\Parity}(x,Y,Z) \equiv
    \neg Z(0) \wedge \forall z{ <} x \; (Z(z+1) \lra (Z(z) \oplus Y(z)))
\]
\begin{defi}
\label{d:V0-2}
\cite{CookNg10}
The theory $V^0(2)$
has vocabulary $\LtwoA$ and axioms those of $V^0$ and
$\exists Z {\le} x{+}1 \; \delta_{\Parity}(x,Y,Z) $.
\end{defi}
The complexity class $FAC^0(2)$ is the $AC^0$ closure of the
function $\Parity(x,Y)$, and in fact the $\Sigma^B_1$-definable
functions of $V^0(2)$ are precisely those in $FAC^0(2)$.

The theory $VTC^0$ for the counting class $TC^0$ is defined
similarly to $V^0(2)$, but now the function $\Parity(x,Y)$ is
replaced by the function $\numones(y,X)$, whose value is the
number of elements (i.e. `ones') of $X$ that are less than $y$.
The axiom for $VTC^0$ is based on a $\Sigma^B_0$ formula
$\delta_\textrm{\it NUM}(y,X,Z)$ defining the graph of a string function
accumulating the values of $\numones(y,X)$ as $y$ increases.

\begin{defi}\label{d:VTC}
\cite{CookNg10}
The theory $VTC^0$ has vocabulary $\LtwoA$ and axioms those of
$V^0$ and $ \exists Z {\le} 1{+}\langle y,y\rangle \;
\delta_\textrm{\it NUM}(y,X,Z)$.
\end{defi}
The class $FTC^0$ is the $AC^0$ closure of the function $\numones$,
and in fact the $\Sigma^B_1$-definable functions of $VTC^0$ are
precisely those in $FTC^0$.

In Chapters 5 and 9 of \cite{CookNg10} it is shown that the theories
$V^0, V^0(2), VTC^0$ have respective universally axiomatized
conservative extensions
$\overline{V^0},\overline{V^0(2)}, \overline{VTC^0}$
obtained by introducing function symbols and their defining
axioms for all string functions
(and some number functions) in the corresponding complexity class.
These have the following properties.

\begin{prop}\label{p:bar}
Let $(FC, V, \overline{V})$ be any of the triples
$(FAC^0, V^0, \overline{V^0})$ or $(FAC^0(2),$
$V^0(2), \overline{V^0(2)})$ or $(FTC^0, VTC^0, \overline{VTC^0})$,
and let $\calL$ be the vocabulary of $\overline{V}$.
\begin{enumerate}[\em(i)]
\item 
$\overline{V}$ is a
universally axiomatized conservative extension of $V$, 
\item
the $\Sigma^B_1$-definable functions of both $V$ and $\overline{V}$
are those in $FC$, 
\item
a string function (respectively number function)
is in $FC$ iff it has a function symbol (respectively term) in
$\calL$,
\item
$\overline{V}$ proves the $\Sigma^B_0(\calL)\ind$
and $\Sigma^B_0(\calL)\comp$ schemes,
\item
for every $\Sigma^B_1(\calL)$
formula $\varphi^+$ there is a $\Sigma^B_1$ formula $\varphi$
such that $\overline{V} \vdash \varphi^+ \lra \varphi$.
\qed
\end{enumerate}
\end{prop}


\section{The New Theories}


When developing theories for $\parityL$ and $\nL$ it will be more
convenient to work first with $\parityL$, where each number can be
represented by a single bit, before adding the complication of
multi-bit numbers in $\nL$.

\subsection{The Theory \texorpdfstring{$V\parityL$}{VparityL}}\label{s:VparityL}

The theory $V\parityL$ is an extension of $V^0(2)$
(Definition \ref{d:V0-2}) obtained by adding an axiom
showing the existence of matrix powering over $\Z_2$.  We use
the fact that $\parityL$ is the $AC^0$ closure of this matrix
powering function (Proposition \ref{p:completeness}) and the
development in Chapter 9 of \cite{CookNg10} to show that the
provably total functions in $V\parityL$ are exactly the set
$F\parityL$.

The actual method followed below of describing the theory $\VpL$ is
slightly more complicated.  In order to obtain the desired axiom for
matrix powering (i.e., the function $\Powtwo$ in Definition \ref{d:Pow}),
we will work with the theory $\overline{V^0(2)}$, a
conservative extension of $V^0(2)$, and its
vocabulary $\calL_{\operatorname{\it FAC}^0(2)}$, which has function
symbols or terms for every function in $\operatorname{\it FAC}^0(2)$
(Proposition \ref{p:bar}).
We describe the $\Sigma^B_1$ axiom for matrix powering
as a $\Sigma^B_1(\calL_{FAC^0(2)})$ formula and refer to part (v) of
proposition \ref{p:bar} to conclude that this formula is provably
equivalent to a $\Sigma^B_1(\LtwoA)$ formula.
Hence we will freely use $\operatorname{\it FAC}^0$ functions and the
function $\Parity(x,Y)$ when describing formulas which will help
express the $\Sigma^B_1$ axiom. 

Each string function $F(\vec{x},\vec{X})$ (other than $\Parity(x,Y))$
in the vocabulary of 
$\overline{V^0(2)}$ has a defining axiom specifying its bit graph
as follows:
\[
 F(\vec{x},\vec{X})(z) \lra
         z< t(\vec{x},\vec{X})\wedge \varphi(z,\vec{x},\vec{X})
\]
where $t$ is an $\LtwoA$ term bounding the function
and the formula $\varphi$ is $\Sigma^B_0$
in the previously defined symbols.

In order for a formula to encode matrix powering, it will need to
include witnesses -- enough information to ``check'' that the matrix
has been powered correctly.  The function $\PowSeqtwo$ computes every
entry of every power of matrix $X$ up to the $k^\textrm{th}$ power.

\begin{defi}[$\PowSeqtwo$] \label{d:PowSeq}
Let $X$ be a string representing an $n \times n$ matrix over $\Z_2$,
and let $X^i$ be the string representing the $i^\textrm{th}$ power of
the same matrix.  Then the string function $\PowSeqtwo(n,k,X)$ has output
the list of matrices $[ID(n), X,X^2,\ldots,X^k]$, coded as described
in Section \ref{s:specialfns}.
\end{defi}

Note that the functions $\Powtwo$ and $\PowSeqtwo$ are $AC^0$-reducible
to each other, so both are complete for $F\parityL$.   Our axiom
for $\VpL$ states the existence of values for $\PowSeqtwo$.

We start by defining the matrix product operation.
The $AC^0$ string function $ID(n)$ codes the $n\times n$ identity
matrix, and has the defining axiom (recall Section \ref{s:specialfns})
\[
ID(n)(b) \lra b< \langle n,n\rangle \wedge
\Pair(b) \wedge \pairleft(b) = \pairright(b)
\]
For $X$ and $Y$ encoding two matrices, the $AC^0$ string function
$G(n,i,j,X,Y)$ codes the string of pairwise bit products of row $i$ of
$X$ and column $j$ of $Y$: 
\[ G(n,i,j,X,Y) (b) \lra b<n\wedge X(i,b)\wedge Y(b,j)\]
We define
\begin{equation}\label{e:PARdef}
\PAR(X) \equiv \Parity(|X|,X)(|X|)
\end{equation}
so that $\PAR(X)$ holds iff
$X$ has an odd number of ones. 
Thus the function $\Prodtwo(n,X,Y)$ producing the product of two
$n\times n$ matrices $X$ and $Y$ over $\Z_2$, has bit graph axiom:
\begin{eqnarray}
\Prodtwo(n,X,Y)(b) & \lra &
b<\langle n,n\rangle \wedge \Pair(b) \wedge \pairleft(b)< n 
\wedge \pairright(b) < n \nonumber \\
&& \wedge \PAR(G(n,\pairleft(b),\pairright(b),X,Y))
\label{e:prodTwo}
\end{eqnarray}
This yields a $\Sigma^B_0(\calL_{FAC^0(2)})$ formula
$\delta_{\PowSeqtwo}(n,k,X,Y)$ for the graph of $\PowSeqtwo(n,k,X,Y)$: 
\begin{eqnarray}\label{e:deltaPowSeqtwo}
& & Y^{[0]} = ID(n) \wedge \forall i{<}k \; (Y^{[i+1]} =
  \Prodtwo(n,X,Y^{[i]}) \label{e:powseqtwo} \\
& & \wedge \forall b {<} |Y| \; \big( Y(b)  \supset
(\Pair(b)\wedge \pairleft(b) < n)\big) \nonumber
\end{eqnarray}
The second line ensures that $Y$ is uniquely defined by specifying
that all bits not used in encoding matrix entries \emph{must} be
false.
Note that $\delta_{\PowSeqtwo}$ involves the function $\Prodtwo$ and
hence is not equivalent to a $\Sigma^B_0$ formula, but by part (v) of
Proposition \ref{p:bar}, $\overline{V^0(2)}$ proves it is equivalent to a
$\Sigma^B_1$ formula $\delta'_{\PowSeqtwo}$.

\begin{defi}\label{d:VParityL}
The theory $V\parityL$ has vocabulary $\LtwoA$ and axioms those
of $V^0(2)$ and the $\Sigma^B_1$ formula
$ \exists Y {\le} 1{+}\langle k,\langle n,n\rangle\rangle \;
    \delta'_{\PowSeqtwo}(n,k,X,Y)$
stating the existence of a string value for the function
$\PowSeqtwo(n,k,X)$. 
\end{defi}

Following the guidelines of Section IX.2 of \cite{CookNg10}, we
prove that several functions, including the aggregate function
$\PowSeqtwo^\star$ (see (\ref{e:PowSeqStarL}) and (\ref{e:PowSeqStar})
below), are $\Sigma_1^B$-definable in $\VpL$.

\begin{lem} \label{l:definableVpL}
The functions $\Powtwo$, $\PowSeqtwo$, and $\PowSeqtwo^\star$ are
$\Sigma_1^B$-definable (definition \ref{d:SigDefine})
in $\VpL$.
\end{lem}

\proof
Since $\overline{V^0(2)}$ is a conservative 
extension of $V^0(2)$, it follows that the theory
\begin{equation}\label{e:calTDef}
\calT=V\parityL + \overline{V^0(2)}
\end{equation}
is a conservative extension of $V\parityL$, and this allows us to
reason in $\calT$ to make inferences about the power of $V\parityL$.
Thus by part (v) of Proposition \ref{p:bar}, to prove that a function
is $\Sigma^B_1$-definable in $\VpL$ it suffices to prove that
it is $\Sigma^B_1(\calL_{FAC^0(2)})$-definable in $\calT$.

For the function $\PowSeqtwo$ the graph is given by either of the
equivalent formulas $\delta_{\PowSeqtwo}$ or $\delta'_{\PowSeqtwo}$.
Existence of the function value follows immediately from the axiom
given in Definition \ref{d:VParityL}, and uniqueness is proved by
induction (justified by part (iv) of Proposition \ref{p:bar})
on the $\Sigma^B_0(\calL_{FAC^0(2)})$
formula stating that the first $i$ bits of the
function's output are unique.

For $\Powtwo$ the function value can be extracted from the value
for $\PowSeqtwo$ by an $AC^0$ function, and existence and uniqueness
of that value follows from existence and uniqueness for $\PowSeqtwo$.

Recall from chapter 8 of \cite{CookNg10} (Definition VIII.1.9) that the
aggregate function $\PowSeqtwostar$ is the polynomially bounded string
function that satisfies 
\begin{equation}\label{e:PowSeqStarL}
 | \PowSeqtwostar(b,W_1,W_2,X)| \leq \langle b, \langle |W_2|,
\langle |W_1|,|W_1|\rangle \rangle \rangle
\end{equation}
and
\begin{equation}\label{e:PowSeqStar}
\PowSeqtwostar(b,W_1, W_2,X)(i,v) \leftrightarrow i<b \wedge
\PowSeqtwo((W_1)^{i}, (W_2)^{i},X^{[i]})(v)
\end{equation} 
The strings $W_1$, $W_2$, and $X$ encode $b$-length lists of inputs to
each place of $\PowSeqtwo$: $W_1$ encodes the list of numbers
$n_0,n_1,n_2,\ldots,n_{b-1}$; $W_2$ encodes the list of numbers
$k_0,k_1,\ldots,k_{b-1}$; and $X$ encodes the list of strings
$X_0,X_1, \ldots, X_{b-1}$.  Our goal is to raise the
$n_i\times n_i$ matrix encoded in $X_i$ to the powers
$1,2,\ldots,k_i$.  The string function $\PowSeqtwostar$ computes
\emph{all} these lists of powers, aggregating many applications of the
function $\PowSeqtwo$.  This is accomplished by forming a large
matrix $S$ by placing padded versions of the matrices
$X_0,X_1, \ldots, X_{b-1}$ down its diagonal, and noting that the
$i$th power of $X_j$ can be extracted from the $i$th power of $S$.
The padding is necessary to make the diagonal matrices all the same
size, so the location of $X_j$ in $S$ can be computed with $AC^0(2)$
functions.
The details of a formula for the graph of $\PowSeqtwo^\star$ and the
proof that $\calT$ can prove its unique output exists
can be found in appendix \ref{a:definableVpL}. 
\qed

Section IX.2 in \cite{CookNg10} presents a general method for defining a
universal conservative extension $\overline{VC}$ (satisfying the
properties of Proposition \ref{p:bar}) of a theory $VC$
over $\LtwoA$.  Here $VC$ is assumed to be defined in a manner similar to
$V^0(2)$ and $VTC^0$; namely by adding a $\Sigma^B_1$ axiom to $V^0$
showing
the existence of a complete function $F_C$ (and its aggregate $F^\star$)
for the complexity class $C$.
Although our new theory $V\parityL$ fits this
pattern, for the purpose of defining $\overline{V\parityL}$ it is
easier to start from the conservative extension 
$\calT$ of $V\parityL$ defined in (\ref{e:calTDef}).

We now view $\calT$ as the extension of  $\overline{V^0(2)}$
obtained by adding the axiom
\[
  \exists Y {\le} 1{+}\langle k,\langle n,n\rangle\rangle \;
    \delta_{\PowSeqtwo}(n,k,X,Y)
\]
where $\delta_{\PowSeqtwo}$ is the original version of
$\delta'_{\PowSeqtwo}$ given in (\ref{e:deltaPowSeqtwo}).
(Note that this axiom is equivalent to the one in Definition
\ref{d:VParityL}.)


A small modification to the development in Section IX.2
generalizes it to the case in which the base theory is an extension $V$ of
$V^0$ rather than just $V^0$ (in this case $V = \calT$).
The construction of $\overline{VC}$
works so that Proposition  \ref{p:bar} holds, where now the
complexity class $C$ is the $AC^0$ closure of $\{\calL,F\}$,
where $\calL$ is the vocabulary of $V$ and $F$ is the function
whose existence follows from the axiom.  In the present case,
this allows us to define $\overline{V\parityL}$ satisfying
Proposition  \ref{p:bar}, where now the complexity class $C$
is the $AC^0$ closure of $\{\calL_{FAC^0(2)}, \PowSeqtwo\}$,
which is same as the $AC^0$ closure of $\{\PowSeqtwo\}$, namely
$\parityL$.

To start this construction we need a quantifier-free axiomatization
of $\calT = V\parityL + \overline{V^0(2)}$.  This consists of the
axioms for $\overline{V^0(2)}$ (which are quantifier-free)
together with a quantifier-free defining axiom for $\PowSeqtwo$.
The formula (\ref{e:deltaPowSeqtwo}) for
$\delta_{\PowSeqtwo}(n,k,X,Y)$ (the graph of $\PowSeqtwo$) has bounded
number quantifiers, but these may be eliminated using functions in
$FAC^0(2)$ (see Section V.6 of \cite{CookNg10}).  Thus
$\delta_{\PowSeqtwo}(n,k,X,Y)$ is provably equivalent to a
quantifier-free formula $\delta{''}_{\PowSeqtwo}(n,k,X,Y)$
over the vocabulary $\calL_{FAC^0(2)}$ for $\overline{V^0(2)}$. 
The required quantifier-free defining axiom is
\[
   Y = \PowSeqtwo(n,k,X,Y) \lra \delta{''}_{\PowSeqtwo}(n,k,X,Y)
\]

As a result of this development we have a version of
Proposition \ref{p:bar} for $\overline{\VpL}$:

\begin{thm} \label{p:bartwo}
  Let $\mathcal{L}_{F\parityL}$ be the vocabulary of $\overline{\VpL}$.
  \begin{enumerate}[\em(i)]
  \item $\overline{\VpL}$ is a universal conservative extension of
    $\VpL$,
  \item the $\Sigma_1^B$-definable functions of both $\VpL$ and
    $\overline{\VpL}$ are those of $F\parityL$,
  \item a string function (respectively, number function) is in
    $F\parityL$ iff it has a function symbol (resp., term) in
    $\mathcal{L}_{F\parityL}$,
  \item $\overline{\VpL}$ proves the
    $\Sigma_0^B(\mathcal{L}_{F\parityL})\ind$ and
    $\Sigma_0^B(\mathcal{L}_{F\parityL})\comp$ schemes, and 
  \item for every $\Sigma_1^B(\mathcal{L}_{F\parityL})$ formula
    $\varphi^+$ there is a $\Sigma_1^B$ formula $\varphi$ such that
    $\overline{\VpL} \vdash \varphi^+ \lra \varphi$. 
\qed
  \end{enumerate}
\end{thm}

\subsection{The Theory \texorpdfstring{$V\#L$}{VnumberL}}\label{s:VnL}

Our theory $\VnL$ is associated with the class $DET = AC^0(\#L)$
(Definition \ref{d:DET}).  Just as $V\parityL$ is an extension of
$V^0(2)$, $\VnL$ is the extension of $VTC^0$ (Definition \ref{d:VTC})
obtained by adding an axiom showing the existence of matrix powering
for matrices with binary integer entries.
We use the fact that $DET$ is the $AC^0$ closure of
this matrix powering function (Proposition \ref{p:completeness})
and the development in Chapter 9 of \cite{CookNg10} to show that
the provably total functions in $V\#L$ are exactly the set $FDET$.

The development of $\VnL$ is similar to that of $\VpL$, but has
extra complications.  Before defining matrix multiplication and
powering we must define multiplication and iterated 
sum for binary integers.  Both of these functions are complete
for the complexity class $TC^0$, so we define $V\#L$ as an extension
of the theory $VTC^0$ (recall that $\VpL$ is defined as an extension
of $V^0(2)$).  To define the required functions we work in the
conservative extension $\overline{VTC^0}$ of $VTC^0$
(just as we worked in the
conservative extension $\overline{V^0(2)}$ of $V^0(2)$ when
developing $\VpL$). 

We represent integers by strings.
For a string $X$ encoding an integer $x$, the first bit $X(0)$ indicates
the sign of $x$: $x< 0$ iff $X(0)$.
The rest of string $X$ is a binary representation of $x$, from least
to most significant bit.
The magnitude of an integer can be extracted with the $FAC^0$
function $\intsize(X)$, which simply deletes the low order bit of
the string $X$.  Thus an integer can be recovered
from its string encoding by computing $(-1)^{X(0)} \cdot \intsize(X)$.

The theory $V^0$ defines binary addition $X+_{\N} Y$ over natural
numbers (Example V.2.5 in \cite{CookNg10}), and $VTC^0$ defines binary
multiplication $X \times_{\N} Y$ over natural numbers
and $\Sum_{\N}(n,X)$ (the sum of the list
$X^{[0]},\ldots,X^{[n-1]}$ of natural numbers)
(Section IX.3.6 in \cite{CookNg10}).
These functions over $\N$ can be used to help
develop quantifier-free formulas
over $\calL_{FTC^0}$ which define the corresponding functions
$+_\Z$, $\times_\Z$, and $\Sum_\Z$ over $\Z$.
For example the function $\times_{\N}$ can be used to define the graph
$R_{\times_\Z}(X,Y,Z)$ by
\[
  \intsize(Z) = \intsize(X) \times_\N \intsize(Y)
  \wedge  ( Z(0) \leftrightarrow (X(0) \oplus Y(0)))
\]
The definition of integer addition $+_\Z$ is more complicated
and requires defining the `borrow' relation for subtraction
as well as using $+_\N$.  The definition of $\Sum_\Z(n,X)$
(iterated integer sum) splits the list $X$ into a list of positive
integers and a list of negative integers, sums each using $\Sum_\N$,
and subtracts the second sum from the first.

Matrices of integers are represented by arrays of strings using the
$\Rowtwo$ function (\ref{eq:rowtwo}).
Integer matrix product is the $TC^0$ function $\ProdZ(n,X,Y)$
defined analogously to formula (\ref{e:prodTwo}) for $\Prodtwo$,
and the graph of
iterated matrix product is defined by the $\Sigma^B_0(\calL_{FTC^0})$
formula $\delta_{\PowSeqZ}(n,k,X,Y)$ analogous to $\delta_{\PowSeqtwo}$
(\ref{e:deltaPowSeqtwo}).  (See \cite{Fontes09}) for more details.)

By part (v) of Theorem \ref{p:bar}, $\overline{VTC^0}$ proves $\delta_{\PowSeqZ}$
is equivalent to a $\Sigma^B_1$ formula $\delta'_{\PowSeqZ}$, which
we use in the following axiom for $V\#L$.

\begin{defi}\label{d:VNumL}
The theory $V\#L$ has vocabulary $\LtwoA$ and axioms those of $VTC^0$
and the $\Sigma^B_1$ formula
$ \exists Y {\le} t \; \delta'_{\PowSeqZ}(n,k,X,Y)$
for a suitable bounding term $t$.
\end{defi}

The next two results are proved similarly to Lemma \ref{l:definableVpL}
and Theorem \ref{p:bartwo}.

\begin{lem} \label{l:definableVnL}
The functions $\Pow_Z$, $\PowSeq_Z$, and $\PowSeq_Z^\star$ are
$\Sigma_1^B$-definable in $\VnL$.
\qed
\end{lem}




\begin{thm} \label{p:barZ}
  Let $\mathcal{L}_{F\nL}$ be the vocabulary of $\overline{\VnL}$.
  \begin{enumerate}[\em(i)]
  \item $\overline{\VnL}$ is a universal conservative extension of
    $\VnL$,
  \item the $\Sigma_1^B$-definable functions of both $\VnL$ and
    $\overline{\VnL}$ are those of $FDET$,
  \item a string function (respectively, number function) is in
    $FDET$ iff it has a function symbol (resp., term) in
    $\mathcal{L}_{F\nL}$,
  \item $\overline{\VnL}$ proves the
    $\Sigma_0^B(\mathcal{L}_{F\nL})\ind$ and
    $\Sigma_0^B(\mathcal{L}_{F\nL})\comp$ schemes, and 
  \item for every $\Sigma_1^B(\mathcal{L}_{F\nL})$ formula
    $\varphi^+$ there is a $\Sigma_1^B$ formula $\varphi$ such that
    $\overline{\VnL} \vdash \varphi^+ \lra \varphi$. 
    \qed
  \end{enumerate}
\end{thm}


\section{Interpretations of \texorpdfstring{\LAP}{LAP}}\label{s: interp}
Having established the theories $\VpL$ and $\VnL$, we are interested
in studying which results from linear algebra are provable in these
theories.  This job is made much easier by taking advantage of
Soltys' theory $\LAP$ \cite{Soltys01, SoltysCo04}, which formalizes
linear algebra over an arbitrary field (or integral domain).
As explained in Section \ref{s:intro}, this theory defines
standard matrix functions such as determinant, adjoint, and
characteristic polynomial, in terms of matrix powering, and
formalizes relative proofs of their properties.
We show that $\VpL$ and $\VnL$ prove these same properties by
interpreting $\LAP$ into each of them.  The two interpretations are
different because the intended semantics are different:  the
underlying rings are respectively $\Z_2$ and $\Z$.

Now Theorems \ref{t:equiv} and \ref{t:hardM} in Section \ref{s:intro}
follow from the fact that
each interpretation translates theorems of $\LAP$ into
theorems in the theory.  Actually there are exceptions to this
preservation in the
case of $\VnL$, because the underlying ring $\Z$ is not a field.
However as explained at the beginning of Section \ref{ss:interpZ},
Theorems \ref{t:equiv} and \ref{t:hardM} do indeed hold for $\VnL$.

\subsection{Defining \texorpdfstring{\LAP}{LAP}} \label{ss:defLAP}
Soltys' theory $\LAP$ (for Linear Algebra with matrix Powering) is a
quantifier-free theory based on Gentzen-style sequents.
It has three sorts: indices (i.e. natural numbers)
(represented by $i$, $j$, $k$), field
elements ($a$, $b$, $c$), and matrices ($A$, $B$, $C$) with
entries from the field.

The language $\mathcal{L}_\LAP$ of $\LAP$ has symbols
\[
\begin{array}{l}
  0_{\text{index}},
1_{\text{index}},
+_{\text{index}},
*_{\text{index}},
\dm_{\text{index}},
\operatorname{div},
\operatorname{rem}, 
\row, \col,  \\
0_{\text{field}},
1_{\text{field}},
+_{\text{field}},
*_{\text{field}},
-_{\text{field}}, {}^{-1},
 \e, \sum,\\
\leq_{\text{index}},
=_{\text{index}},
=_{\text{field}},
=_{\text{matrix}},
\text{cond}_{\text{index}},
\text{cond}_{\text{field}},
\p
\end{array}
\]

The intended meanings of $0, 1, +, *, ^{-1}$, and $-_{\text{field}}$
are obvious.
The symbol $\dm_{\text{index}}$ is cutoff subtraction;
$\operatorname{div}(i,j)$ and $\operatorname{rem}(i,j)$ are the
quotient and remainder functions;
$\row(A)$ and $\col(A)$ return the numbers of rows and columns in $A$; 
$\e(A,i,j)$ is the field element $A_{ij}$, (where $A_{ij} = 0$ if 
$i=0$ or $j=0$ or $i> r(A)$ or $j> c(A)$),
$\sum(A)$ is the sum of all the entries of $A$;
and for $\alpha$ a formula,
${\operatorname{cond}_{\text{index}}}(\alpha,i,j)$ is $i$ if
$\alpha$ is true and $j$ otherwise (similarly 
for ${\operatorname{cond}_{\text{field}}}(\alpha,a,b)$).
The powering function $\p(n,A) = A^n$.

Terms and quantifier-free formulas are mostly constructed in the usual
way, respecting types.
We use $n,m$ for index terms, $t,u$ for field terms,
$T,U$ for matrix terms, and $\alpha,\beta$ for (quantifier-free)
formulas.  The four kinds of atomic formulas are
$m\le_{\text{index}} n$, $m=_{\text{index}} n$, $t =_{\text{field}} u$,
and $T =_{\text{matrix}} U$.  Formulas are built from atomic formulas
using $\wedge,\vee,\neg$.
There are restrictions on terms beginning with
$\operatorname{cond}$:  If $\alpha$ is a formula with atomic
subformulas all of the form $m\leq_{\text{index}}n$ and
$m=_{\text{index}}n$, then
${\operatorname{cond}}_{\text{index}}(\alpha,m',n')$
is a term of type index and
${\operatorname{cond}}_{\text{field}}(\alpha,t,u)$ is a term of type field.

A term of type matrix is either a variable ($A,B,C,\cdots$) or
a lambda term of the form 
$\lambda_{ij}\langle m,n,t \rangle$ (with the restriction that
$i$ and $j$ are not free in the index terms $m$ and $n$).
This lambda term defines an $m \times n$ matrix with $(i,j)^\textrm{th}$
entry given by $t(i,j)$.

Lines in an $\LAP$ proof are Gentzen-style sequents
$ \alpha_1,\ldots, \alpha_k \ra \beta_1,\ldots,\beta_\ell$
with the usual meaning
$\bigwedge \alpha_i \supset \bigvee \beta_j$.
The logical axioms and rules are those of Gentzen's system LK
(minus the quantifier rules).  The nonlogical axioms are numbered
{\bf A1} through {\bf A36} (see Appendix \ref{a:LAP_Axioms}).

There are two nonlogical rules.
The first nonlogical rule is the Induction Rule:
\begin{equation}\label{e:indRule}
   \frac{\Gamma,\alpha(i) \ra \alpha(i+1),\Delta}
{\Gamma,\alpha(0)\ra \alpha(n),\Delta}
\end{equation}
The second nonlogical rule is the Matrix Equality Rule,
which states that two matrices are equal if they have the same
numbers of rows and columns, and have equal entries:
\begin{equation}\label{e:equalRule}
 \frac{S_1 \ \ S_2 \ \  S_3}{\Gamma\ra\Delta, T=_{\text{matrix}}U}
 \qquad  \mbox{where} \ \
\begin{array}{l}
S_1: \ \  \Gamma\ra\Delta,e(T,i,j)=e(U,i,j) \\
S_2: \ \ \Gamma\ra\Delta,r(T)=r(U) \\
S_3: \ \ \Gamma\ra\Delta,c(T)=c(U)
\end{array}
\end{equation}

Many matrix functions such as multiplication, addition,
transpose, can be defined using $\lambda$ terms, avoiding the need
for separately defined function symbols.
We use the following abbreviations for defined terms.

\medskip
\noindent
{\bf Integer maximum}   $\qquad \max\{i,j\}:=\text{cond}(i\leq j,j,i)$

\medskip
\noindent
{\bf Matrix sum} $ \qquad A+B:=\lambda ij\langle\max\{\row(A),\row(B)\},
            \max\{\col(A),\col(B)\}, A_{ij}+B_{ij}\rangle$

\medskip
Note that $A+B$ is well defined even if $A$ and $B$ are incompatible
in size, because of the convention that out-of-bound entries are 0.

\medskip
\noindent
{\bf Scalar product}
$\qquad aA:=\lambda ij\langle\row(A),\col(A),a*A_{ij}\rangle$

\medskip
\noindent
{\bf Matrix transpose}
$\qquad A^t:=\lambda ij\langle\col(A),\row(A),A_{ji}\rangle$

\medskip
\noindent
{\bf Zero and Identity matrices}\\
$\qquad 
0_{kl}:=\lambda ij\langle k,l,0\rangle \qquad \text{and} \qquad
I_k:=\lambda ij\langle k,k,\text{cond}(i=j,1,0)\rangle$

\medskip
Sometimes we
will just write $0$ and $I$ when the sizes are clear from the context.

\medskip
\noindent
{\bf Matrix trace}
$\qquad \text{tr}(A):=\Sigma\lambda ij\langle\row(A),1,A_{ii}\rangle$

\medskip
\noindent
{\bf Dot product}
$\qquad A\cdot B:=\Sigma\lambda ij\langle
\max\{\row(A),\row(B)\},\max\{\col(A),\col(B)\},
A_{ij}*B_{ij}\rangle$

\medskip
\noindent
{\bf Matrix product}$\\
A*B:=\lambda ij\langle\row(A),\col(B),
\lambda kl\langle\col(A),1,\entry(A,i,k)\rangle
\cdot
\lambda kl\langle\row(B),1,\entry(B,k,j)\rangle\rangle$
\\

Matrix product is defined even when $\col(A) \ne \row(B)$, again
by the convention that out-of-bound entries are 0.

Finally, the following
decomposition of an $n\times n$ matrix $A$
is used in the axioms defining $\Sigma(S)$ and in presenting
Berkowitz's algorithm:
\begin{equation}\label{decomp}
A=\left(\begin{array}{cc}
a_{11} & R \\ S & M
\end{array}\right)
\end{equation}
where $a_{11}$ is the $(1,1)$ entry of $A$, and $R,S$ are
$1\times(n\!-\!1)$, $(n\!-\!1)\times 1$ submatrices, respectively, and
$M$ is the principal submatrix of $A$
Therefore, we make the following precise definitions:
\begin{equation}\label{eq:RSM}
\begin{split}
R(A)&:=\lambda ij\langle 1,\col(A)-1,\entry(A,1,i+1)\rangle \\
S(A)&:=\lambda ij\langle \row(A)-1,1,\entry(A,i+1,1)\rangle \\
M(A)&:=\lambda ij\langle \row(A)-1,\col(A)-1,\entry(A,i+1,j+1)\rangle
\end{split}
\end{equation}\label{RSM}

\subsection{Interpreting \texorpdfstring{$\LAP$}{LAP} into
  \texorpdfstring{$V\parityL$}{VparityL}} \label{ss:interp2}

Here we take the underlying field in the semantics of $\LAP$ to
be $\Z_2$.  We interpret the three-sorted theory $\LAP$
into the two-sorted theory $V\parityL$.  The index sort is interpreted
as the number sort, field elements are interpreted as Boolean values,
and matrices are interpreted as strings.  We translate each
formula $\alpha$ of $\LAP$ into a formula $\alpha^\sigma$
of $\overline{V\parityL}$.  Here $\alpha^\sigma$ is in
$\Sigma^B_0(\calL_{F\parityL})$, so by part(v) of Theorem
\ref{p:bartwo}, $\alpha^\sigma$ is equivalent 
to a $\Sigma^B_1$ formula $(\alpha^{\sigma})'$ of $V\parityL$. 
The translation preserves provability (sequent theorems are translated to
sequent theorems) and it also preserves truth in our intended
standard models: ($\N,\Z_2, \mbox{matrices over $\Z_2$}$) for $\LAP$ and
($\N, \mbox{finite subsets of $\N$}$) for $V\parityL$.

In order for Theorems \ref{t:equiv} and \ref{t:hardM}  (for the case
of $V\parityL$) to follow from this interpretation we need that
a term $T$ of type matrix in $\LAP$, when interpreted as a matrix
over $\Z_2$, is translated to a string term
$T^\sigma$ in $\overline{V\parityL}$ which represents the same
matrix, using conventions like those developed in Sections
\ref{s:specialfns} and \ref{s:VparityL}.  In particular the
matrix entries (elements of $\Z_2$) are certain bits in the
string $T^\sigma$, which is why we represent the elements $0,1$ of
$\Z_2$ as the Boolean values $\bot$ and $\top$.

Recall that formulas of $\LAP$ are built from atomic formulas
using the Boolean connectives $\wedge,\vee,\neg$.  Below we show
how to translate atomic formulas.  Then each formula $\alpha$
is translated by translating its atomic formulas and putting them
together with the same connectives.  Thus
$(\alpha \wedge \beta)^\sigma = \alpha^\sigma \wedge \beta^\sigma$,
$(\alpha \vee \beta)^\sigma = \alpha^\sigma \vee \beta^\sigma$,
and $(\neg \alpha)^\sigma = \neg \alpha^\sigma$.
Finally, a sequent of $\LAP$ is translated to a sequent of
$\overline{\VpL}$ formula by formula.  Thus the translation of
$ \alpha_1,\ldots, \alpha_k \ra \beta_1,\ldots,\beta_\ell$
is $ \alpha_1^\sigma,\ldots, \alpha_k^\sigma \ra
\beta_1^\sigma,\ldots,\beta_\ell^\sigma$

The atomic formulas of $\LAP$ are simple, since the only predicate
symbols are $\le$ and the three sorts of =.  Thus the main work
in defining the interpretation comes in translating the three
sorts of terms: $m$ to $m^\sigma$, $t$ to $t^\sigma$, and $T$ to 
$T^\sigma$.  In order to do this
we define function symbols in the language of $\overline{V\parityL}$
to interpret functions in $\LAP$: $\dm$,
$f_\textrm{div}$, $f_\textrm{rem}$, $f_\textrm{r}$, $f_\textrm{c}$,
and functions $f_\varphi$ and $F_{\varphi}$ for certain formulas
$\varphi$.   All of these functions have $\Sigma^B_0$ definitions and
are definable in the subtheory
$\overline{V^0}$ of $\overline{\VpL}$.  However to interpret
$\Sigma$ and {\sc p} we need to go beyond $\overline{V^0}$.

Defining $\Sigma$ requires the parity function $PAR(X)$,
which is definable in the theory
$V^0(2)$.  We note that the theory $\LAP$ is
the extension of the base theory $\LA$ obtained
by adding the function {\sc p} and the two axioms {\bf A35} and
{\bf A36} defining matrix powering.  Our interpretation translates
$\LA$ into $V^0(2)$.  Conveniently
$\VpL$ is obtained from $V^0(2)$ by adding the axiom defining
$\PowSeqtwo$ (defining matrix powering).

The next three subsections give inductive definitions for translating
$\LAP$ terms of each of the three sorts.

\subsubsection{Index sort}\label{s:indexSort}

The table below shows how to translate each index term $m$ of $\LAP$
into a term $m^\sigma$ in $\overline{\VpL}$, except we postpone
to Subsection \ref{s:matrixSort} translating terms involving
$\row$ and $\col$, which specify the numbers of rows and columns in
a matrix.
The graph of each function
symbol used in $m^\sigma$ is defined by a $\Sigma^B_0$ formula
(except for the case $\operatorname{cond}_\text{index} (\alpha,m,n)$,
for which the formula is in $\Sigma^B_0(\alpha^\sigma))$.  These formulas
are given in the table below.
(The first three lines give the base case of the inductive
definition of the translation.)

\[
\begin{array}{ccl}
\LAP & \overline{\VpL} \\
\hline
0_\text{index} & 0 \\
1_\text{index} & 1 \\
i & i & \text{index variables map to number variables} \\
m+_\text{index} n & m^\sigma + n^\sigma \\
m *_\text{index} n & m^\sigma \cdot n^\sigma \\
m \dm_\text{index} n & m^\sigma \dm n^\sigma &
\text{`$\dm$' is standard limited subtraction, defined:}\\
  & & x \dm y = z \leftrightarrow  (x = y+z) \vee (z=0 \wedge x<y) \\
\operatorname{div}(m,n) & f_\textrm{div}(m^\sigma,n^\sigma) & \text{$f_\textrm{div}$ is a number function with graph:}\\
  & & f_\textrm{div}(x,y)=z \leftrightarrow (y \cdot z \leq x \wedge x < y \cdot (z+1)) \\
  & & \vee (y=0\wedge z=0)\\
\operatorname{rem}(m,n) & f_\textrm{rem}(m^\sigma,n^\sigma) & \text{$f_\textrm{rem}$ is a number function with graph:}\\
  & & f_\textrm{rem}(x,y)=z \leftrightarrow
z + y\cdot f_\textrm{div}(x,y) = x\\
\operatorname{cond}_\text{index} (\alpha,m,n) 
  & f_{\alpha^\sigma}(m^\sigma,n^\sigma) & \text{where for certain
  formulas $\varphi$,
 $f_\varphi$ is defined:}\\
  & & f_\varphi(x,y)=z \leftrightarrow (\varphi \wedge x=z) \vee (\neg \varphi \wedge y=z)
   \end{array} \]
Note that in $\LAP$, $\operatorname{cond}_\text{index}(\alpha,m,n)$ is
defined only when all atomic sub-formulas of $\alpha$ have the form
$m=n$ or $m\leq n$.  Thus $f_\varphi(x,y)$ need only be defined when
$\varphi$ is the interpretation of such a formula from $\LAP$.

\subsubsection{Field sort}\label{s:fieldSort}

Since we represent elements of $\Z_2$ by
Boolean values in $V\parityL$, each field term $t$ is translated to a
$\Sigma^B_0(\calL_{F\parityL})$ formula $t^\sigma$.  (This convenient
representation of ring elements as truth values is not possible for
the ring $\Z$, complicating section \ref{ss:interpZ}.)
Each variable of type field is interpreted as a formula specifying
the first entry of a $1\times 1$ matrix variable.

The table below shows how to interpret variables of type field
(except those involving the matrix entry function e and the matrix
sum function $\Sigma$, which are given in Subsection \ref{s:matrixSort}).

For $t$ and $u$ terms of type field, we interpret:
\[
\begin{array}{ccl} 
\LAP &  \VpL \\
\hline
0_\text{field} & \bot 
  \\
1_\text{field} & \top
  \\
a & X_a(1,1) & \text{$X_a$ is a string variable} \\
t +_\text{field} u & t^\sigma \oplus u^\sigma 
   & \text{that is, $t^\sigma\operatorname{XOR}u^\sigma$}\\
t-_\text{field}u & t^\sigma \oplus u^\sigma 
  \\
t *_\text{field} u & t^\sigma \wedge u^\sigma 
  \\
t^{-1} & t^\sigma 
  \\
\operatorname{cond}_\text{field}(\alpha, t,u) & (\alpha^\sigma \wedge
t^\sigma) \vee (\neg \alpha^\sigma \wedge u^\sigma) 
\end{array}
\]

\subsubsection{Matrix sort}\label{s:matrixSort}

The translation of terms involving matrices is complicated.  Every matrix
of $\LAP$ has three attributes:  number of rows, number of columns,
and matrix entries (field elements).  In our two-sorted language
for $\VpL$ we represent a matrix by a string which codes all of these. 
Thus an $a\times b$
matrix $A$ is represented by a string $A^\sigma$ such that 
$A^\sigma(0,\langle a,b\rangle)$ is true, and for all $i,j$
with $1\le i \le a$ and $1 \le j \le b$ and $e(A,i,j) = A_{ij} =1$,
the bit $A^\sigma(i,j)$ is true.  All other bits of $A^\sigma$ are false.

We will use the formula $\isMatrixtwo(X)$ (equivalent to a $\Sigma^B_0$
formula)
which asserts that the string $X$ properly encodes a matrix as above.
We allow the number of rows and/or columns to be 0,
but any entry out of bounds is 0 (a false bit).
\begin{eqnarray}\label{e:isMat}
& & \isMatrixtwo(X) \  \equiv  \
\exists x,y{<} |X| \; \ \big[ X(0, \langle x,y \rangle) \wedge  \\
& & \forall z {<} |X| \;  \big((z=0\vee\neg  X(z,0)) \wedge  
 (z=\langle x,y\rangle \vee \neg X(0,z))  \big) \wedge  \nonumber \\
& &  [ X(z) \supset \big(\Pair(z) \wedge  
\pairleft(z)\leq x\wedge
(\pairleft(z)=0 \vee \pairright(z) \leq y) \big)] \big] \nonumber
\end{eqnarray}

The following table completes the inductive definition of the
translation of all $\LAP$ terms.  (Here $\operatorname{\it PAR}(X)$ is 
defined in (\ref{e:PARdef}) and the functions $f_r,f_c,F_\varphi,F_p$
are defined below).)

\[
\begin{array}{cll}
\LAP & \overline{\VpL} \\
\hline
A & A  \qquad  \text{matrix variables map to string variables}\\
\row(T) & f_r(T^\sigma) \\
\col(T) & f_c(T^\sigma) \\
\e(T,m,n) & T^\sigma(m^\sigma,n^\sigma)
\wedge m^\sigma > 0 \wedge n^\sigma > 0
\wedge \isMatrixtwo(T^\sigma)  \\
\sum(T) & \neg \operatorname{\it PAR}(T^\sigma) \wedge
           \isMatrixtwo(T^\sigma)  \\
\lambda_{ij}\langle m,n,t\rangle & F_{t^\sigma}(m^\sigma,n^\sigma) \\
\p(m,T) & F_p(m^\sigma,T^\sigma)  
\end{array}
\]

The above translation is designed so that for every $\LAP$ matrix
term $T$ (except a matrix variable $A$), $\isMatrixtwo(T^\sigma)$
holds and is provable in $\overline(\VpL)$.

The $\overline{V\parityL}$ functions $f_r(X)$ and $f_c(X)$ extract
the numbers of rows and columns of the matrix coded by $X$, and
are used  in the table above to translate the $\LAP$ terms
$\row(T)$ and $\col(T)$.  These have defining equations
\[ f_r(X)=z \lra  (\neg \isMatrixtwo(X) \wedge z=0) \vee
    \big(\isMatrixtwo(X) \wedge \exists y {\le} |X| \; X(0, \langle z,y
          \rangle) \big)  \]
\[ f_c(X)=z \lra (\neg \isMatrixtwo(X) \wedge z=0) \vee
    \big(\isMatrixtwo(X) \wedge \exists y {\le} |X|  \; X(0, \langle y,z
          \rangle) \big) \]
Note that strings not encoding a matrix are semantically interpreted
as the $0\times 0$ matrix.

The translation of the matrix entry term $\ent(T,m,n)$ is
consistent with the $\LAP$ convention that rows and columns are
numbered starting with 1.  A string not properly encoding a
matrix has no non-zero entries.

The translation of the entry sum term $\sum(T)$ is $\bot$ if $T^\sigma$
does not properly encode a matrix, and otherwise it is the
parity of 1 + the number of one-bits in $T^\sigma$ (the extra
bit is $T^\sigma(0,\langle a,b\rangle)$ coding the number of
rows and columns).

The matrix term $\lambda_{ij}\langle m,n,t\rangle$ is interpreted
by the $\overline{V\parityL}$ term $F_{t^\sigma}(m^\sigma,n^\sigma)$.
Here $F_{t^\sigma}(x,y)$ is a string function, which has additional
arguments corresponding to any free variables in $t^\sigma$
other than the distinguished variables $i,j$ (we interpret $i^\sigma =i$
and $j^\sigma =j$).  The bit defining
formula for $F_{t^\sigma}$ is
\[
   F_{t^\sigma}(x,y)(b) \lra b = \langle 0,\langle x,y\rangle \rangle
\vee \exists i{\le} x \; \exists j{\le} y \;
  (i>0\wedge j>0 \wedge
b = \langle i,j\rangle \wedge t^\sigma(i,j))
\]
where $t^\sigma(i,j)$ indicates the distinguished
variables $i,j$.
Then $\isMatrixtwo(F_{t^\sigma}(m^\sigma,n^\sigma))$
is always true (and provable in $\overline{\VpL)}$).

The last line in the table above translates the $\LAP$ term
$\p(m,T)$ representing matrix powering to the
$\overline{\VpL}$ term $ F_p(m^\sigma,T^\sigma)$.  The function
$F_p$ is defined in terms of $\Powtwo$ (Definition \ref{d:Pow})
which defines matrix powering in $\VpL$.  But the definition of $F_p$
is not straightforward, because our translation of $\LAP$ terms
includes the extra bit in row 0 coding the matrix dimension.

Thus we need two $\Sigma_0^B$ string functions.  The first one strips
the dimensions from the string, converting a matrix encoded according
to our interpretation (of $\LAP$ into $\VpL$) into a matrix according
to the standard in $\VpL$.
\[ \strip(X)(i,j) \leftrightarrow X(i+1,j+1) \]
The second function adds the dimension-encoding ``wrapper'' back,
converting a standard-form $\VpL$ matrix into the form of an
interpreted matrix from $\LAP$.
\[ \wrap(r,c,X)(b) \leftrightarrow b=\langle 0, \langle r,c\rangle\rangle \vee \exists 0 {<} i,j {<} b \; (b=\langle i,j \rangle \wedge X(i,j)) \]
Given these two functions, we want $F_p(i,X)$ to interpret the $\LAP$
function $\p(i,X)$.  
Let $r_X=f_r(X)$ be the interpretation of $\operatorname{r}(X)$,
and $c_X = f_c(X)$ be the interpretation of $\operatorname{c}(X)$.
For strings $X$ that satisfy $\isMatrixtwo(X)$ and powers $i>0$,
 \[ F_p(i,X) = \wrap(r_X, c_X, \Powtwo (\max(r_X,c_X), i, \strip(X))) \]
This is consistent with the $\LAP$ convention that for $i>1$,
$\p(i,A)$ retains
the row and column dimensions of $A$, and is defined by raising the
matrix $A'$ to the power $i$ and truncating excess rows or columns, where
$A'$ is the square matrix of dimension $\max\{\row(A),\col(A)\}$
obtained from $A$ by adding 0 entries where needed.

There is a special case: if $i=0$, then $\LAP$ specifies that the
zeroth power of a matrix $A$ is the $\operatorname{r}(A) \times
\operatorname{r}(A)$ identity matrix. 
 \[ F_p(0,X) = \wrap (r_X, r_X, \Powtwo(r_X, 0, \strip(X))) \]
If $\neg \isMatrixtwo(X)$ then $F_p(i,X)$ codes the $0\times 0$ matrix.

Combining all these cases, we can bit-define $F_p$ as follows:
\[
\begin{array}{l}
F_p(i,X)(b)  \leftrightarrow \\
\big[ \isMatrixtwo(X) \wedge i=0 \wedge 
    \wrap (r_X, r_X, \Powtwo(r_X, 0, \strip(X)))\big]\\
 \vee \big[ i>0 \wedge
\wrap(r_X, c_X, \Powtwo (\max(r_X,c_X), i, \strip(X)))\big] \\
\vee \big[\neg\isMatrixtwo(X)\wedge
        b=\langle 0,\langle 0,0\rangle\rangle\big]
        \label{interpret-pow}
\end{array}
\]

\subsubsection{Translating atomic formulas}\label{s:atomic}
We translate atomic formulas as follows:

\[\begin{array}{cl}
\LAP & \overline{\VpL}  \\
\hline
m=_\text{index} n & m^\sigma =_1 n^\sigma \\
m \leq_\text{index} n & m^\sigma \leq n^\sigma\\
t=_\text{field} u & t^\sigma \leftrightarrow u^\sigma\\
T=_{\text{matrix}} U & (\row(T) = \row(U))^\sigma \wedge
(\col(T) = \col(U))^\sigma \wedge \\
&   \forall i,j \leq (|T^\sigma|+|U^\sigma|) \
(\e(i,j,T) = \e(i,j,U) )^\sigma

\end{array}
\]
The translation of $T=_{\text{matrix}} U$ could be simplified to
$T^\sigma = U^\sigma$ for all $\LAP$ terms $T$ and $U$ other than
matrix variables $A$, since all other terms translate to terms
satisfying $\isMatrixtwo$.    However two distinct string variables
$A,B$ could nevertheless represent matrices that are equal
according to the Matrix Equality Rule (\ref{e:equalRule}) if
they do not satisfy $\isMatrixtwo$.

\subsubsection{Provability is preserved}\label{s:provability}

\begin{thm}\label{t:provability}
If $\LAP$ proves $ \alpha_1,\ldots, \alpha_k \ra \beta_1,\ldots,\beta_\ell$
then $\overline{\VpL}$ proves \\
$ \alpha_1^\sigma, \ldots, \alpha_k^\sigma~\ra~
\beta_1^\sigma,\ldots,\beta_\ell^\sigma$.
\end{thm}

\proof
(For more details see \cite{Fontes10}.)
Since our translation of formulas and sequents is transparent to
the logical connectives, the logical axioms and rule applications
for $\LAP$ (the propositional part of Gentzen's system LK)
translate to logical axioms and rule applications for
$\overline{\VpL}$.

The nonlogical axioms {\bf A1},...,{\bf A36} for $\LAP$ are given in
Appendix \ref{a:LAP_Axioms}
and the nonlogical rules are (\ref{e:indRule}) (Induction) and
(\ref{e:equalRule}) (Matrix Equality).
It suffices to show that the translated axioms
are theorems of $\overline{\VpL}$, and for each of the rules, if the
translated hypotheses are theorems of $\overline{\VpL}$ then
so is the translated conclusion.

The two rules are easily handled.  Consider
the Induction Rule (\ref{e:indRule}).
Since the translation of each sequent of $\LAP$ is a sequent
of quantifier-free formulas in $\overline{\VpL}$, the fact that
this translated rule preserves theorems in $\overline{\VpL}$
follows from the fact that $\overline{\VpL}$ proves the induction
scheme for $\Sigma_0^B(\mathcal{L}_{F\parityL})$ formulas
(part (iv) of Theorem \ref{p:bartwo}).

Now consider the Matrix Equality Rule (\ref{e:equalRule}).
The fact that this translated rule
preserves theorems in $\overline{\VpL}$ follows immediately 
from the way that matrix equality $=_\text{matrix}$ is translated
(Section \ref{s:atomic}).

Now consider the axioms of $\LAP$.
Axioms {\bf A1} to {\bf A5} are equality axioms, and their translations
mostly follow from the usual equality axioms.
However $=_\text{matrix}$ is not translated as equality, but it
does translate to an equivalence relation,
and the functions which take
a matrix $A$ as an argument (namely r, c, e, $\sum$, and {\sc p})
depend only on the numbers of rows and columns of $A$, and the
entries of $A$.  Hence all translations of {\bf A1} to {\bf A5}
are theorems of $\overline{\VpL}$.

Axioms {\bf A6} to {\bf A14} translate to simple properties of
$\N$ under $+,\cdot$, and $\le$, which are provable in $V^0$
(see Chapter III of \cite{CookNg10}).  {\bf A15}, {\bf A16}, and
{\bf A17} translate to simple properties of $\N$ which are
easily proved in $\overline{V^0}$
from the definitions $f_\text{div}, f_\text{rem}$, and $f_\varphi$.

Axioms {\bf A18} to {\bf A26} are the axioms that define a field.
Section \ref{s:fieldSort} translates field terms to formulas
(where $\bot$ represents 0 and $\top$ represents 1).   All these
axioms translate into logical tautologies, so they are all trivially
provable in $\overline{\VpL}$.  For example
\[
\begin{array}{ll}
{\bf A18}: \ \  \ra 0\ne 1 \wedge a+0=a & \text{translates to}\\
{\bf A18}^\sigma: \ (\bot \lra \neg\top)\wedge
((X_a(1,1)\oplus\bot)\lra X_a(1,1))
\end{array}
\]

Axiom {\bf A27} defines
cond$_\text{field}$, and its instances also translate into tautologies.

Axioms {\bf A28} and {\bf A29} define the row, column, and entry
functions r, c, e, and relate them to the lambda terms defining
matrices.  The translations of these follow easily from the definitions
of $f_r, f_c, F_\varphi$, and are provable in $\overline{V^0}$.

Axioms {\bf A30} to {\bf A34} define $\sum(A)$ recursively by
breaking the matrix $A$ into four parts as illustrated in (\ref{decomp}).
Since the translation of $\sum(A)$ is defined in terms of the
parity function $PAR$, the following lemma is useful.
It states that if two strings differ on exactly one bit,
then they have opposite parities.
\begin{lem} \label{l:oneBit}
  $\overline{V^0(2)}$ proves:
  \[
  \begin{array}{l}
    (X(k)\leftrightarrow \neg Y(k)) \wedge \\
    \big( \forall i< |X| + |Y| \; i \neq k \leftrightarrow
    (X(i)\leftrightarrow Y(i)) \big) \supset 
    \big( \operatorname{\it PAR}(X) \leftrightarrow \neg
    \operatorname{\it PAR}(Y) \big) 
  \end{array}
  \]
\end{lem}

\proof
To prove the lemma, recall (\ref{e:PARdef}) that
$\PAR(X) \equiv \Parity(|X|,X)(|X|)$.
The proof in $\overline{V^0(2)}$ proceeds by induction on the bits of the
witness strings $\Parity(|X|,X)$ and $\Parity(|Y|,Y)$
computing the parities of $X$ and $Y$.
\qed

To prove the translations of axioms {\bf A28} and {\bf A29}
in $\overline{V^0(2)}$ we consider separately
the two cases $\neg\isMatrixtwo(A)$ and $\isMatrixtwo(A)$.
The first case is trivial to prove because then $A$ has zero rows
and columns and it has no entries.  So we may assume $\isMatrixtwo(A)$.
In that case, according to Section \ref{s:matrixSort}, $\sum(A)$
translates to $\neg \PAR(A)$.

Axiom {\bf A30} asserts $\sum(A) = \ent(A,1,1)$ in case $A$ has
exactly one row and column.  Since we may assume $\isMatrixtwo(A)$,
the string $A$ has at most two 1-bits: one is $A(0,\langle 1,1\rangle)$
specifying its dimension, and the other possibility is $A(1,1)$.
With the help of Lemma \ref{l:oneBit}, $\overline{V^0(2)}$ proves
$A(1,1) \lra \neg \PAR(A)$, as required.

Axiom {\bf A31} states that if $A$ has exactly one row and at least
two columns, then $\sum(A) = \sum(B) + A_{1\col(A)}$, where
$B$ is the matrix obtained from $A$ by deleting the last entry in row 1.
The string $Y$ representing $B$ differs in at most three bits from
the string $X$ representing $A$.   Two  of these are in row 0, since
the dimensions of $A$ and $B$ are different, and the third possibility
is $X(1,\col(A))$ versus $Y(1,\col(A))$, where the latter is always 0.
Hence $\overline{V^0(2)}$ proves the translation of the axiom using
three applications of Lemma \ref{l:oneBit}.

Axiom {\bf A32} states that if $A$ has only one column, then
$\sum(A)=\sum(A^t)$.  (See Section \ref{ss:defLAP} for the
definition of $A^t$, the transpose of $A$.)  The translation is
equivalent to
\[
    f_c(A)=1\wedge \isMatrixtwo(A) \supset (\PAR(A) \lra \PAR(B))
\]
where $B = F_\varphi(1,f_r(A))$ is the term translating $A^t$.
The more general statement obtained by replacing the 
conclusion $\PAR(A) \lra \PAR(B))$ by asserting roughly  that the parity of
the first $i$ rows of $A$ equals the parity of the first $i$ columns
of $B$ by induction on $i$, using Lemma \ref{l:oneBit}.
The precise statement takes into account the bit in row 0 of
each matrix describing the dimension of the matrix.

Axiom {\bf A33} asserts that if $A$ has at least two rows and
two columns, then $\sum(A)$ is the the sum of the four submatrices
pictured in (\ref{decomp}).   The translation asserts that the
parity of $A$ is equivalent to the exclusive or of the parities
of the four pieces.  This is proved by double induction, first
by considering just the first $i$ rows of both sides, and for
the induction step, to consider in addition the first $j$ columns
of row $i+1$.

Axiom {\bf A34} assert $\sum(A) = 0$ in case $A$ has 0 rows or
0 columns.  Its translation is easily proved.

Axiom {\bf A35} is $\ra \text{\sc p}(0,A)=I_{\row(A)}$.
The term $\text{\sc p}(0,A)$ is translated $F_p(0,A)$, and 
the term $I_{\row(A)}$ is translated $F_{t^\sigma}(f_r(A),f_r(A))$,
where $t$ is the term cond$(i=j,1,0)$ (see the definition of
$I_k$ in Section \ref{ss:defLAP}).  The symbol = in the axiom
is really $=_{\text{matrix}}$, so the translated axiom asserts
that the number of rows, number of columns, and entries of the
translations of the two matrix terms, are all respectively equal. 
Proving this
amounts to verifying that all the defined functions have their intended
meanings.

Axiom {\bf A36} is $\ra \text{\sc p}(n+1,A)=\text{\sc p}(n,A)\ast A$.
The proof of the translation distinguishes the two cases
$\neg\isMatrixtwo(A)$ and $\isMatrixtwo(A)$.  The former is easy,
since both sides of the equation translate into the zero by zero
matrix.  $\overline{\VpL}$ proves the second case by induction
on $n$.   Recall that {\sc p}$(n,A)$ is translated $F_p(n,A)$,
where $F_p$ is defined in terms of $\Powtwo$, which in turn is
defined in terms of $\PowSeqtwo$ (\ref{e:deltaPowSeqtwo}),
whose value is a sequence of 
successive powers of its matrix argument.  The induction proof
involves verifying that the translated lambda term defining matrix
product in Section \ref{ss:defLAP} correctly corresponds to the
definition (\ref{e:prodTwo}) defining $\Prodtwo$ in $\VpL$.
\qed

\subsection{Interpreting \texorpdfstring{$\LAP$}{LAP} into
  \texorpdfstring{$V\#L$}{VnumberL}}\label{ss:interpZ} 

Now we interpret the underlying `field' in the semantics of $\LAP$
to be the ring $\Z$ of integers.  Of course $\Z$ is not a
field, so we cannot translate the axiom {\bf A21}
$a \ne 0 \ra a \ast (a^{-1}) =1$ for field inverses.
However according to the footnote on page 283 of \cite{SoltysCo04},
this axiom is not used except in the proof of Lemma 3.1 and Theorem 4.1.
We do not need Lemma 3.1, and it is not hard to see that Theorem 4.1
does hold for integral domains (the field inverse axiom can be
replaced
by the axiom $a\ast b = 0, a \ne 0 \ra b=0$ prohibiting zero divisors).
It follows that $\LAP$ over integral domains proves the hard
matrix identities in Definition \ref{d:hardM} and the equivalence
of (i), (ii), and (iii) in Section \ref{s:intro}.
Thus our interpretation of $\LAP$ into $V\#L$ will show
that $V\#L$ proves both of these results, and so Theorems \ref{t:equiv}
and \ref{t:hardM} hold for $V\#L$.

The interpretation of $\LAP$ into $V\#L$ is similar to the
interpretation into $\VpL$.  We first translate $\LAP$ into
$\overline{V\#L}$, and then into $V\#L$, using
part (v) of Theorem \ref{p:barZ}. 
As before, it suffices to show how to translate terms and atomic
formulas, since general formulas and sequents are translated by
translating their atomic formulas.
We do not translate terms or formulas involving the field inverse
function $t^{-1}$.

We use $t^\sigma$ and $\alpha^\sigma$ to denote the translation
of $\LAP$ terms $t$ and formulas $\alpha$.

Terms of type index (which do not involve matrices)
are translated into number terms in exactly
the same way as for $\VpL$ (see Section \ref{s:indexSort}).  

Terms of type field
(integers) are now translated into strings representing integers
in binary, using the functions $+_\Z$ and $\times_\Z$ described
in Section \ref{s:VnL}. 

For $t$ and $u$ terms of type field in $\LAP$ we interpret:

\[\begin{array}{ccl}
\LAP &  \VnL  \\
\hline
0_\text{field} & \text{the empty string } \varnothing
  \\
1_\text{field} & \text{the string ``$10$'',} & \text{ i.e., $X$ such that } X(i)
\leftrightarrow i=1
  \\
a & X_a & \text{field variables map to string variables} \\
t +_\text{field} u & t^\sigma +_\Z u^\sigma 
   \\
t-_\text{field}u & t^\sigma +_\Z (u^\sigma)' 
        & \text{ where $(u^\sigma)'$ is identical to $u^\sigma$,
except} \\
& & \text{on the first bit} \\
t *_\text{field} u & t^\sigma \times_\Z u^\sigma  
  \\
\operatorname{cond}_\text{field}(\alpha, t,u) & F^\text{cond}_{\alpha^\sigma}
(t^\sigma, u^\sigma) 
        & \text{ where for each formula $\varphi$, define: }\\
        && F^\text{cond}_\varphi (X,Y)=Z  \leftrightarrow \\
       && (\varphi \wedge X=Z) \vee (\neg \varphi \wedge Y=Z)
\end{array}\]

$\LAP$ terms $T$ of type matrix are translated into string terms
$T^\sigma$ representing arrays of binary integers.    As in the
translation into $\VpL$,  if $T$ has $r$ rows and $c$ columns, then
$T^\sigma(0,x)$ holds iff $x = \langle r,c\rangle$.  For $1\le i\le r$
and $1\le j \le c$, the binary integer entry $T_{ij}$ is given
by $Row_2(i,j,T^\sigma)$, where $Row_2$ is defined in (\ref{eq:rowtwo}).
We need a formula $\isMatrixz(X)$ asserting that $X$ is a valid
encoding of a matrix of integers.  This is defined similarly to
$\isMatrixtwo$ for the translation into $\VpL$.

The following table shows how to translate terms involving matrices.
\[\begin{array}{cll}
\LAP & \overline{\VnL} \\
\hline
A & A  \qquad  \text{matrix variables map to string variables}\\
\row(T) & f^\#_r(T^\sigma) \\
\col(T) & f^\#_c(T^\sigma) \\
\e(T,m,n) & F_e(m^\sigma,n^\sigma,T^\sigma) \\
\sum(T) & F_{\sum}(T^\sigma)  \\
\lambda_{ij}\langle m,n,t\rangle & F^\#_{t^\sigma}(m^\sigma,n^\sigma) \\
\p(m,T) & F^\#_p(m^\sigma,T^\sigma)  
\end{array}
\]
The functions above with superscripts \# are similar to their
prototypes for the translation into $\VpL$.
Thus $f^\#_r$ and $f^\#_c$ have the same definitions as $f_r$ and
$f_c$ except $\isMatrixtwo$ is replaced by $\isMatrixz$.
The entry function $F_e$ is defined by
\[
   F_e(i,j,X)(b) \lra X(i,(j,b))\wedge i>0\wedge j>0 \wedge \isMatrixz(X)
\]
The function $F_{\sum}(X)$ sums the integer entries of the matrix $X$,
and is defined in terms of the iterated integer sum function 
$Sum_\Z(n,Y)$ mentioned in Section \ref{s:VnL} (where $Y$ is a
rearrangement of the integer entries of the matrix $X$ to form a linear
list).

The matrix term $\lambda_{ij}\langle m,n,t\rangle$ is interpreted by
the $\VnL$ term $F^\#_{t^\sigma}(m^\sigma,n^\sigma)$.   Here
$F^\#_{t^\sigma}(x,y)$ has additional arguments corresponding to any
free variables in $t^\sigma$ other than the distinguished variables
$i,j$.  Its bit defining axiom is similar to that of $F_t^\sigma$,
used to translate lambda terms into $\VpL$.
\[
\begin{array}{l}
F^\#_{t^\sigma}(x,y)(b) \lra \\
b = \langle 0,\langle x,y\rangle\rangle \vee \exists i {\le} x \;
\exists j {\le} y \; \exists k {\le} b \; \big(i>0\wedge j>0\wedge
b=\langle i,\langle j,k\rangle\rangle \wedge t^\sigma(i,j)(k)\big)
\end{array}
\]
The  function $F^\#_p$ interpreting matrix powering is defined similarly
to $F_p$ for the $\VpL$ case, except that the functions $\wrap$ and
$\strip$ need to be modified, $\Powtwo$ is replaced by $\Pow_Z$,
and $\isMatrixtwo$ is replaced by $\isMatrixz$.

To complete the definition of the interpretation, we note that
atomic formulas are translated as in Section \ref{s:atomic}.

\subsubsection{Provability is preserved}\label{s:proveNumber}

\begin{thm}\label{t:NumProvability}
If $\LAP$ proves $ \alpha_1,\ldots, \alpha_k \ra
\beta_1,\ldots,\beta_\ell$ and none of the formulas $\alpha_i$
or $\beta_j$ contains a term of the form $t^{-1}$,
then $\overline{\VnL}$ proves
$ \alpha_1^\sigma,\ldots, \alpha_k^\sigma \ra
\beta_1^\sigma,\ldots,\beta_\ell^\sigma$,
where now $\sigma$ refers to the translation of $\LAP$
into $\overline{\VnL}$.
\end{thm}

\proof
As in the case for $\VpL$, the proof involves showing that the
translated axioms (with the axiom prohibiting zero divisors replacing the
field inverse axiom {\bf A21}) are theorems of $\overline{\VnL}$,
and the translated rules preserve provability in $\overline{\VnL}$.
(The proof for each axiom and rule is similar to the proof for the case
of $\VpL$, except for the field axioms, which we discuss below.)
This shows that the translation $S^\sigma$ of a theorem $S$ of $\LAP$
is provable in $\overline{\VnL}$, provided that the proof of $S$
in $\LAP$ does not use the field inverse axiom {\bf A21}.  However
Theorem \ref{t:NumProvability}
makes a stronger claim, namely that the proof of $S$ can
use the original axiom {\bf A21}, provided that no term $t^{-1}$ involving
field inverses occurs in $S$.  This holds because the theorems $S$
are quantifier-free (semantically the free variables are universally
quantified).  Reasoning model-theoretically, if $S$ holds for an
arbitrary field, then it also holds for an arbitrary integral domain,
because the domain can be extended to a field (the field of fractions).

The only $\LAP$ axioms whose translations in $\VnL$ require proof methods
significantly different than for $\VpL$ are the field axioms
{\bf A18} to {\bf A26} (where we replace {\bf A21}
$a\neq 0\ra a*(a^{-1})=1$ by the integral domain axiom
$a*b=0, a\ne 0 \ra b=0$).   Showing that $\overline{\VnL}$
(in fact $\overline{VTC^0}$) proves
the commutative, associative, and distributive laws for the operations
 $+_\Z$ and $\times_Z$ over the binary integers is tedious.
This is not much different than proving that the operations $+_N$ and
$\times_N$ over binary natural numbers satisfy these laws.
Some proof outlines and hints for the latter can be found in
Section IX.3.6 of \cite{CookNg10}.
\qed


\section{Conclusion}\label{s:conclusion}

There are two general motivations for associating theories with
complexity classes.  The first is that of reverse mathematics
\cite{Simpson}:
determining the complexity of concepts needed to prove various
theorems, and in particular whether the correctness of an algorithm
can be proved with concepts of complexity comparable to that of
the algorithm.  The second motivation comes from propositional
proof complexity:  determining the proof lengths of various 
tautology families in various proof systems.
To explain these we start by stating the following:

\begin{oprob}\label{o:quest}
Can $V\parityL$ prove the Cayley-Hamilton Theorem or the
`hard matrix identities' (Definition \ref{d:hardM}) over $\Z_2$?
Can $V\#L$ prove these over $\Z$?
\end{oprob}

The importance of these questions stems partly from Theorem \ref{t:equiv},
which states that the theories prove the equivalence of
the Cayley-Hamilton Theorem and two other properties of the determinant,
and from Theorem \ref{t:hardM}, which states that the theories prove
that the Cayley-Hamilton Theorem implies the `hard matrix identities'.
Theorems \ref{t:equiv} and \ref{t:hardM} follow from the
corresponding theorems in $\LAP$ (for which Open Questions \ref{o:quest}
also apply), and from our interpretations (Sections \ref{s:provability}
and \ref{s:proveNumber}).

It is possible that these questions (Open Questions \ref{o:quest})
could be answered positively without answering the corresponding
questions for $\LAP$, by using methods not available to $\LAP$.  
For example a proof in $V\parityL$
might be able to take advantage of the simplicity of $\Z_2$, or
a proof in $V\#L$
might be able use the algorithmic strength of integer matrix
powering (as opposed to matrix powering over an unspecified field)
to prove correctness of the dynamic programming
algorithm for the determinant in \cite{MV97}.   This algorithm
is based on a combinatorial characterization of $det(A)$
using clow (closed walk) sequences in the edge-labeled graph
specified by the matrix $A$.

Over the field $\Z_2$
the hard matrix identities translate naturally to a family of
propositional tautologies (and over $\Z$ they translate into
another family of tautologies).   The motivation
for studying these identities is to give further examples
of tautology families (like those in \cite{BBP94})
that might be hard for the class of propositional proof
systems known as Frege systems.   There is a close connection
between the strength of a theory needed to prove these identities
(or any $\Sigma^B_0$ formula) and the strength of the propositional
proof system required for their propositional translations to
have polynomial size proofs.  (Chapter 10 of \cite{CookNg10} gives
propositional proof systems corresponding in this way
to five of the theories in (\ref{e:mesh}).)

In particular, the fact that the hard matrix identities
are provable in $VP$ shows that their propositional translations
have polynomial size
proofs in Extended Frege systems.  If the identities were provable in
$VNC^1$ then the tautologies would have polynomial size Frege
proofs.  If the identities turn out to be provable in one
of our new theories, then the tautologies would have polynomial
size proofs in proof systems (yet to be defined) of strength
intermediate between Frege and Extended Frege systems.

Finally, we point out a lesser open problem.
The main axiom for our new theory $V\#L$ asserts that
integer matrix powers exist, where integers are represented in
binary.  As explained at the end of Section \ref{s:newClasses}
integer matrix powering is complete for the complexity class
$DET$ even when restricted to 0-1 matrices, because the binary case
is $AC^0$-reducible to the 0-1 case.  It would be interesting
to investigate whether the nontrivial reduction (see \cite{Fontes09})
can be proved correct in the base theory $VTC^0$, so that $V\#L$ could
equivalently be axiomatized by the axiom for the 0-1 case rather than
the binary case.


\pagebreak
\begin{center} {\huge \bf Appendix} \end{center}
\appendix

\section{Details of Lemma \ref{l:definableVpL}} \label{a:definableVpL}

Recall that in order to show that $\PowSeqtwo^\star$ is
$\Sigma_0^B(\mathcal{L}_A^2)$-definable in $\VpL$, we need to show both:
\[
  \PowSeqtwostar(b,W_1,W_2,X) 
  = Y \leftrightarrow \delta_{\PowSeqtwostar}(b,W_1,W_2,X,Y) 
\]
and
\[V\parityL \vdash \forall b \forall X, W_1, W_2 \exists ! Y
\delta_{\PowSeqtwostar}(b,W_1,W_2,X,Y) \]

We introduce the $AC^0$ functions $\operatorname{\it max}$ and $S$ in
order to simplify the definition of $\delta_{\PowSeqtwostar}$ over
$\overline{V^0(2)}$, and then use proposition \ref{p:bar}
to obtain a provably equivalent $\Sigma_1^B(\mathcal{L}_A^2)$ formula.
Since $\overline{V^0(2)}$ is a conservative extension of $V^0(2)$,
this suffices to show that $\PowSeqtwostar$ is
$\Sigma_1^B(\mathcal{L}_A^2)$-definable in $V\parityL$.  In order to
do so, we must establish a $\Sigma_1^B(\mathcal{L}_{FAC^0(2)})$
formula equivalent to the desired formula $\delta_{\PowSeqtwostar}$.

Let the function $\operatorname{\it max}(n,W)$ yield the maximum
number from a list of $n$ numbers encoded in string $W$:
\[
  \operatorname{\it max}(n,W)=x \leftrightarrow \exists i<n \forall j<n, x=
  (W)^i \geq
  (W)^j \label{eq:max_def}
\]

The string function $S$ can be bit-defined as follows.  Consider two
strings $W_1$, representing a list of $b$ numbers, and $X$,
representing a list of $b$ matrices as above.  The function
$S(b,W_1,X)$ returns the matrix with matrices $X_i$ (appropriately
padded with zeroes) on the diagonal, and all other entries zero.  Let
$m = \operatorname{\it max}(b,W_1)$.  Let $X_i'$ be the $m\times m$
matrix $X_i$ padded with columns and rows of zeroes:
  \[ \begin{array}{rc}
    X_i & \overbrace{\begin{array}{ccc} 0 & \ldots & 0 \end{array}}^{m-n_i} \\
    m-n_i \left \{ \begin{array}{c} 0 \\ \vdots \\ 0 \end{array} \right . &
    \begin{matrix} \ddots & & \vdots \\  &\ddots \\ \ldots && 0
    \end{matrix}
  \end{array} \]
Then $S(b,Y_1,X)$ is the string encoding the matrix:
  \[ \begin{bmatrix}
    \begin{array}{cc}
      X_0' \\ & X_2'
    \end{array}
    & 0 \\
    0 &
    \begin{array}{cc}
      \ddots \\ & X_{b-1}'
    \end{array}
  \end{bmatrix} \]
All entries not in the matrices along the diagonal are $0$.

The string function $S$ can be bit-defined:
  \begin{eqnarray*}
    S(b,W_1,X)(i,j) &\leftrightarrow&
    \exists a<b, \exists i', j' < (W_1)^a, 
    i < m \wedge j< m \wedge \\
    &&     i=i' + m \wedge j=j'+m \wedge X^{[a]}(i',j')
  \end{eqnarray*}
By convention, the unspecified bits (i.e., bits $b$ that are not pair
numbers) are all zero.  This bit-definition ensures that the string
$S(b,W_1,X))$ is uniquely defined.

We will use this matrix $S(b,W_1,X)$ and the existence and uniqueness
of its sequence of matrix powers to show the existence and uniqueness
of the aggregate matrix powering function.

Let $n_{\max}$ denote $\operatorname{\it max}(b,W_1)$ and $k_{\max}$
denote $\operatorname{\it max}(b,W_2)$.
Recall equation (\ref{e:powseqtwo}) defining
$\delta_{\PowSeqtwo}(n,k,X)$.  By convention, the aggregate function 
of $\PowSeqtwo$ is defined as:
  \begin{eqnarray}
    \PowSeqtwostar(b,W_1,W_2,X) = Y  \leftrightarrow 
    |Y| < \langle b, \langle k_{\max},\langle n_{\max},n_{\max}\rangle \rangle
\rangle
    \wedge 
    \hspace{1in} \nonumber \\ 
    \forall j < |Y|, \forall i < b,
    \big[ (Y(j) \supset \Pair(j)) \wedge 
    \delta_{\PowSeqtwo}((W_1)^i,(W_2)^i,X^{[i]},Y^{[i]}) \big] 
  \label{formula:implicit-PowSeq2*}
  \end{eqnarray}

The right-hand side of (\ref{formula:implicit-PowSeq2*}) is the
relation $\delta_{\PowSeqtwostar}(b,W_1,W_2,X,Y)$; it has a provably
equivalent $\Sigma_1^B(\mathcal{L}_A^2)$-formula $\exists Z<t,
\alpha_{\PowSeqtwostar}(b,W_1,W_2,X,Y,Z)$, used as the definition for
$\PowSeqtwostar$ over $V\parityL$.

It remains to prove the existence and uniqueness for
$\PowSeqtwostar(b,W_1,W_2,X)$.  The functions $\operatorname{\it max}$
and $S$ allow for a straightforward proof based on the existence and
uniqueness of the string $A=\PowSeqtwo(b\cdot n_{\max}, k_{\max},
S(b,W_1,X))$, where $n_{\max} = \operatorname{\it max}(b,W_1)$ and
$k_{\max} = \operatorname{\it max}(b,W_2)$.

We would like to define the string $B = \PowSeqtwostar(b,W_1,W_2,X)$
from $A$.  Notice that $B$ encodes a list of strings, each of which
represents a power of the matrix $S(b,W_1,X)$.  The string $A$ encodes
nearly the same information, but in a different format: $A$ is a list
of \emph{lists}, each of which encodes the powers of a matrix from the
list $X$ of matrices.

Observe that:
\[ S(b,W_1,X)^i = 
\begin{bmatrix}
    \begin{array}{cc}
      X_0' \\ & X_2'
    \end{array}
    & 0 \\
    0 &
    \begin{array}{cc}
      \ddots \\ & X_{b-1}'
    \end{array}
  \end{bmatrix}^i
= 
\begin{bmatrix}
    \begin{array}{cc}
      X_0'^i \\ & X_2'^i
    \end{array}
    & 0 \\
    0 &
    \begin{array}{cc}
      \ddots \\ & X_{b-1}'^i
    \end{array}
  \end{bmatrix} \]
Also, 
\[X_j'^i = 
\begin{array}{rc}
    X_j^i & \overbrace{\begin{array}{ccc} 0 & \ldots &
        0 \end{array}}^{m-n_j} \\ 
    m-n_j \left \{ \begin{array}{c} 0 \\ \vdots \\ 0 \end{array}
    \right . & 
    \begin{matrix} \ddots & & \vdots \\  &\ddots \\ \ldots && 0
    \end{matrix}
  \end{array} \]
Thus we can ``look up" the required powers of each matrix.
Let $A$ and $B$ be shorthand:
\[A=\PowSeqtwo( b\cdot n_{\max}, k_{\max}, S(b,W_1,X))\]
\[B = \PowSeqtwostar (b,W_1,W_2,X)\]
Then we can define $\PowSeqtwostar$ from $\PowSeqtwo$ as follows.
\begin{eqnarray}
\hspace{-1cm}{B^{[m][p]}}(i,j) &\leftrightarrow&
m < b \wedge p \leq (W_2)^m \wedge i < (W_1)^m \wedge j<(W_1)^m \wedge
\nonumber 
\\
&&\exists m'<m, m'+1=m \wedge
A^{[p]}(n_{\max}\cdot m'+i, n_{\max}\cdot m'+j)
\label{eq:PS2star-ABdef}
\end{eqnarray}
Here, $m$ represents the number of the matrix in the list $X$, $p$
represents the power of matrix $X_m$, and $i$ and $j$ represent the
row and column; thus the formula above defines the bit $(X_m^p)(i,j)$
for all matrices $X_m$ in the list $X$.  By shifting around the pieces
of this formula and adding quantifiers for $m$, $p$, $i$, and $j$, it
is clear that (\ref{eq:PS2star-ABdef}) can be translated into the
appropriate form for a $\Sigma_1^B$-definition of the graph of
$\PowSeqtwostar$.

Thus existence and uniqueness of $\PowSeqtwostar(b,W_1,W_2,X)$ follow
from existence and uniqueness of $\PowSeqtwo(b\cdot
n_{\max},k_{\max},S(b,W_1,X)$.  Since $\operatorname{\it max}$ and $S$ are $AC^0$
functions, they can be used without increasing the complexity of the
definition (again, by use of proposition \ref{p:bar}, as above).

\section{Axioms for \texorpdfstring{$\LAP$}{LAP}}\label{a:LAP_Axioms}

Here we use the abbreviations given at the end of Section
\ref{ss:defLAP}, such as $A+B, A^t$ etc.  We use $A_{ij}$ for
$e(A,i,j)$.

\noindent
{\bf Equality Axioms}

These are the usual equality axioms, generalized to apply to the
three-sorted theory LA.
Here = can be any of the three equality symbols, $x,y,z$ are variables
of any of the three sorts (as long as the formulas are syntactically
correct).  In {\bf A4}, the symbol $f$ can be any of the nonconstant
function
symbols of $\LAP$.  However {\bf A5} applies only to $\leq$,
since this in the only predicate symbol of $\LAP$ other than =.

\smallskip

\noindent
{\bf A1} $\ra x=x$\\
{\bf A2} $x=y\ra y=x$\\
{\bf A3} $(x=y\wedge y=z)\ra x=z$\\
{\bf A4} $x_1=y_1,...,x_n=y_n\ra fx_1...x_n=fy_1...y_n$\\
{\bf A5} $i_1=j_1,i_2=j_2,i_1\leq i_2\ra j_1\leq j_2$
\smallskip\begin{center}
{\bf Axioms for indices}
\end{center}\smallskip
{\bf A6} $\ra i+1\not= 0$\\
{\bf A7} $\ra i*(j+1)=(i*j)+i$\\
{\bf A8} $i+1=j+1\ra i=j$\\
{\bf A9} $\ra i\leq i+j$\\
{\bf A10} $\ra i+0=i$\\
{\bf A11} $\ra i\leq j, j\leq i$\\
{\bf A12} $\ra i+(j+1)=(i+j)+1$\\
{\bf A13} $i\leq j, j\leq i\ra i=j$\\
{\bf A14} $\ra i*0=0$\\
{\bf A15} $i\leq j, i+k=j\ra j \dm i=k$ \ \textbf{and}
\ $i\nleq j\ra j \dm i=0$\\
{\bf A16} $j\neq 0\ra \text{rem}(i,j)<j$ \ \textbf{and} \
$j\neq 0\ra i=j*\text{div}(i,j)+\text{rem}(i,j)$\\
{\bf A17} $\alpha\ra\text{cond}(\alpha,i,j)=i$ \ \textbf{and} \
$\neg\alpha\ra\text{cond}(\alpha,i,j)=j$
\smallskip\begin{center}
{\bf Axioms for field elements}
\end{center}
{\bf A18} $\ra 0\not= 1 \wedge a+0=a$\\
{\bf A19} $\ra a+(-a)=0$\\
{\bf A20} $\ra 1*a=a$\\
{\bf A21}\footnote{{\bf A21} can be replaced by $a*b=0, a\ne 0 \ra b=0$
for the purpose of proving Theorems \ref{t:equiv} and \ref{t:hardM}.}
$a\neq 0\ra a*(a^{-1})=1$\\
{\bf A22} $\ra a+b=b+a$\\
{\bf A23} $\ra a*b=b*a$\\
{\bf A24} $\ra a+(b+c)=(a+b)+c$\\
{\bf A25} $\ra a*(b*c)=(a*b)*c$\\
{\bf A26} $\ra a*(b+c)=a*b+a*c$\\
{\bf A27} $\alpha\ra\text{cond}(\alpha,a,b)=a$ \ \textbf{and} \
$\neg\alpha\ra\text{cond}(\alpha,a,b)=b$
\smallskip\begin{center}
{\bf Axioms for matrices}
\end{center}\smallskip
Axiom {\bf A28} states that
$\ent(A,i,j)$ is zero when $i,j$ are outside the size of $A$.  Axiom
{\bf A29} defines the behavior of constructed matrices.
Axioms {\bf A30-A33} define the function
$\Sigma$ recursively by first defining it for row vectors,
then column vectors (recall $A^t$  is the transpose of $A$),
and then in general using the decomposition (\ref{eq:RSM}).
Finally, axiom {\bf A34} takes care of empty matrices.

\smallskip
\noindent
{\bf A28} $(i=0\vee\row(A)<i\vee j=0\vee\col(A)<j)\ra\ent(A,i,j)=0$\\
{\bf A29} $\ra\row(\lambda ij\langle m,n,t\rangle)=m$ \ \textbf{and} \
$\ra\col(\lambda ij\langle m,n,t\rangle)=n$ \ \textbf{and} \\
$1\leq i, i\leq m, 1\leq j, j\leq n
\ra\ent(\lambda ij\langle m,n,t\rangle,i,j)=t$\\
{\bf A30} $\row(A)=1,\col(A)=1\ra\Sigma(A)=\ent(A,1,1)$\\
{\bf A31} $\row(A)=1, 1<\col(A)\ra
\Sigma(A)=\Sigma(\lambda ij\langle 1,\col(A)-1,A_{ij}\rangle)
+A_{1\col(A)}$\\
{\bf A32} $\col(A)=1\ra\Sigma(A)=\Sigma(A^t)$\\
{\bf A33} $1<\row(A), 1<\col(A) \ra \Sigma(A)=\ent(A,1,1)
+\Sigma(R(A))+\Sigma(S(A))+\Sigma(M(A))$ \\
{\bf A34} $\row(A)=0\vee\col(A)=0\ra\Sigma A=0$

\smallskip\begin{center}
{\bf Axioms for matrix powering}
\end{center}\smallskip

\noindent
{\bf A35}\footnote{This version is from page 45 of \cite{Soltys01}}
$\ra \text{\sc p}(0,A)=I_{\row(A)}$ \\
{\bf A36}  $\ra \text{\sc p}(n+1,A)=\text{\sc p}(n,A)\ast A$.

\pagebreak
\bibliographystyle{alpha}
\bibliography{sources}

\end{document}